\documentclass{article}

\usepackage{arxiv}

\usepackage[utf8]{inputenc} 
\usepackage[T1]{fontenc}    
\usepackage{hyperref}       
\usepackage{url}            
\usepackage{booktabs}       
\usepackage{amsfonts}       
\usepackage{nicefrac}       
\usepackage{microtype}      
\usepackage{lipsum}		
\usepackage{graphicx}
\usepackage{doi}
\usepackage{amsmath}
\usepackage{lineno}
\usepackage{xcolor}
\usepackage{siunitx}
\usepackage{tabularx}
\usepackage{array}
\newcolumntype{L}[1]{>{\raggedright\arraybackslash}p{#1}}
\newcolumntype{C}[1]{>{\centering\arraybackslash}p{#1}}
\newcolumntype{R}[1]{>{\raggedleft\arraybackslash}p{#1}}
\usepackage{soul}
\usepackage{todonotes}
\usepackage{array, makecell}

\usepackage{ragged2e}

\title{Photonic neural networks based on integrated silicon microresonators}

\author{ \href{https://orcid.org/0000-0002-3361-133X}{\includegraphics[scale=0.06]{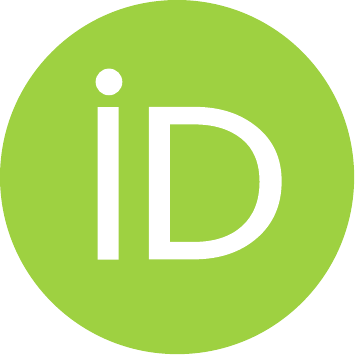}\hspace{1mm}Stefano Biasi, } 
	\And
	\href{https://orcid.org/0000-0001-6156-8394}{\includegraphics[scale=0.06]{orcid.pdf}\hspace{1mm}Giovanni Donati,} 
 	\And
	\href{https://orcid.org/0000-0002-6587-2614}{\includegraphics[scale=0.06]{orcid.pdf}\hspace{1mm}Alessio Lugnan,} 
 \And
	\href{https://orcid.org/0000-0002-8335-1847}{\includegraphics[scale=0.06]{orcid.pdf}\hspace{1mm}Mattia Mancinelli,}
 \And
	\href{https://orcid.org/0000-0002-4167-9158}{\includegraphics[scale=0.06]{orcid.pdf}\hspace{1mm}Emiliano Staffoli,}
  \And
	\href{https://orcid.org/0000-0001-7316-6034}{\includegraphics[scale=0.06]{orcid.pdf}\hspace{1mm}Lorenzo Pavesi}
  	 \\ \\
    Nanoscience Laboratory, Department of Physics, University of Trento, Italy. 
    \\
    Corresponding authors: stefano.biasi@unitn.it, lorenzo.pavesi@unitn.it 
    \\
}

\renewcommand{\shorttitle}{\textit{arXiv} Template}

\hypersetup{
pdftitle={A template for the arxiv style},
pdfsubject={q-bio.NC, q-bio.QM},
pdfauthor={David S.~Hippocampus, Elias D.~Striatum},
pdfkeywords={First keyword, Second keyword, More},
colorlinks=true,
}

\begin{document}
\maketitle

\begin{abstract}
The recent progress of artificial intelligence (AI) has boosted the computational possibilities in fields where standard computers are not able to perform. The AI paradigm is to emulate human intelligence and therefore breaks the familiar architecture on which digital computers are based. In particular, neuromorphic computing, artificial neural networks (ANN) and deep learning models mimic how the brain computes. Large networks of interconnected neurons whose synapsis are individually strengthened or weakened during the learning phase find many applications. With this respect, photonics is a suitable platform to implement ANN hardware thanks to its speed, low power dissipation and multi-wavelength opportunities. One photonic device candidate to perform as an optical neuron is the optical microring resonator. Indeed microring resonators show both a nonlinear response and a capability of optical energy storage, which can be interpreted as a fading memory. Moreover, by using silicon photonics, the photonic integrated circuits can be fabricated in volume and with integrated electronics on board. For these reasons, here, we describe the physics of silicon microring resonators and of arrays of microring resonators for application in neuromorphic computing. We describe different types of ANNs from feed-forward networks to photonics extreme learning machines and reservoir computing. In addition, we discuss also hybrid systems where silicon microresonators are coupled to other active materials. this review aims to introduce the basics and to discuss the most recent developments in the field.
\end{abstract}


\section{Introduction}
Artificial Neural Networks (ANNs), where data is processed in a way that is inspired by the human brain \cite{schuman2022opportunities}, have shown unprecedented computation capabilities \cite{zhang2021study} at the expense of long training times and huge power consumption \cite{mcdonald2022great}. ANNs are becoming increasingly popular due to their versatility in solving a wide range of problems \cite{Genty2020ultrafast}. Large ANNs outperform human minds in certain tasks \cite{meta2022human,li2022competition}. ANNs are usually implemented on electronic hardware which are based on Von Neuman architectures, such as general purpose CPUs (Central Processing Units), massively parallel GPUs (Graphical Processing Units), and specialized integrated circuits dedicated to accelerate specific operations like TPUs (Tensor Processing Units) \cite{sze2017efficient,bhattacharya2021dnns,dhilleswararao2022efficient}. Electronic ANNs face challenges such as long training times and high power consumption  \cite{strubell2019energy,wu2022sustainable,boahen2022dendrocentric} as well as signal interference, difficulty in handling floating point operations, and low parallel computing efficiency \cite{Sui2020review,Liu2021research,Porte2021complete}.

\begin{figure}[h!]
	\centering
	\includegraphics[width=0.8\textwidth]{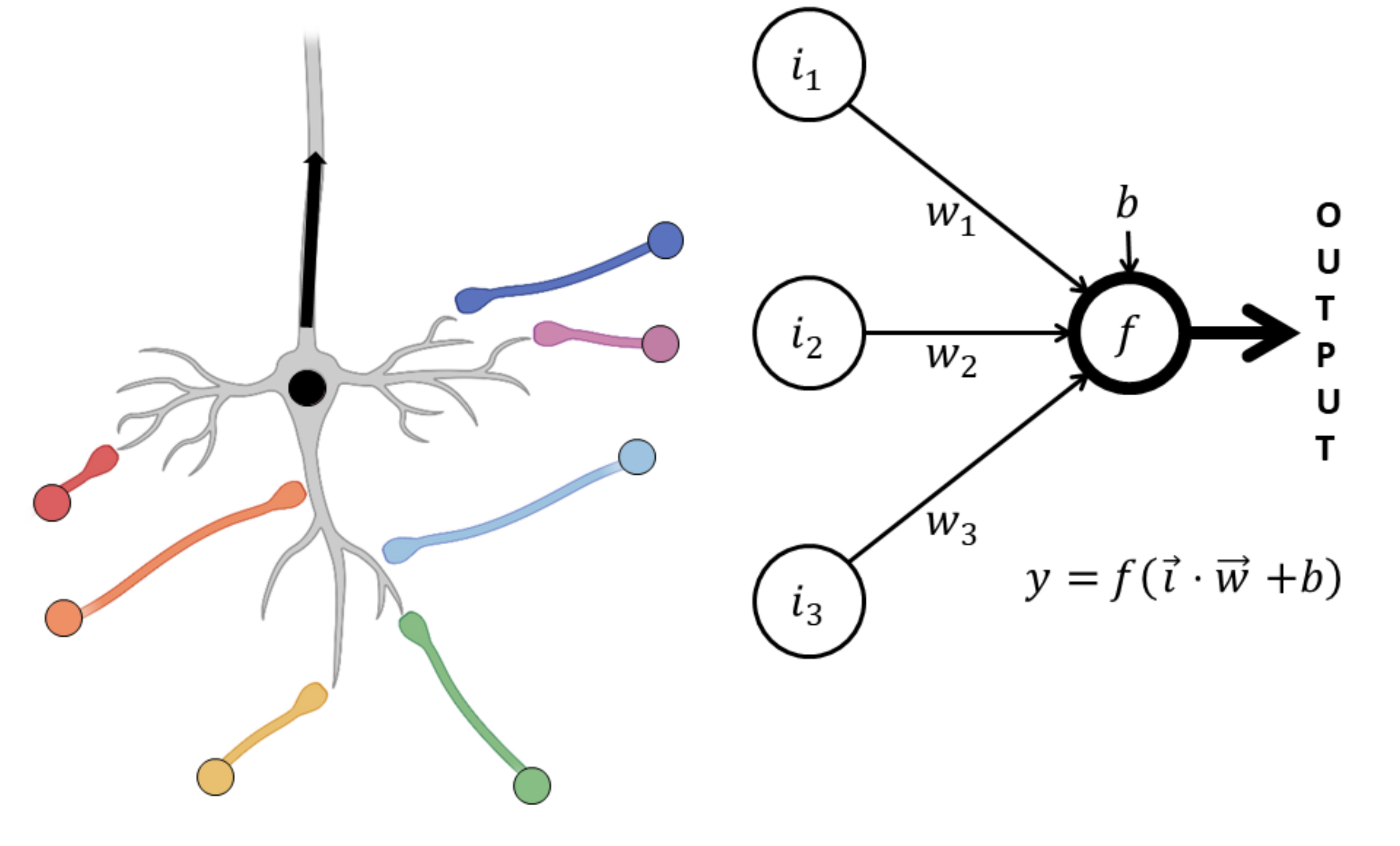}
	\caption{Left: Sketch of a biological neuron (black dot). Different signals from nearby neurons (colored) are collected by the dendrites through interconnecting synapses. The neuronal body integrates the signals and, if above a threshold, produces a voltage spike which is sent via the axons (black arrow) to the post-synaptic neurons. Right: Sketch of an artificial neuron where the output $y$ is produced by the formula given in the inset from the different inputs $i$ (image courtesy of Gianmarco Zanardi).}
	\label{fig:opticalNeuron}
\end{figure}

Most of the issues associated to electronic hardware arises from the massive amount of data that has to be moved among the different parts of the circuits. Photonics enables low loss and low latency interconnects, where data throughput and parallelism can be greatly enhanced via wavelength division multiplexing. Therefore, ANN can be significantly accelerated and improved by implementing the required connections in a photonic hardware. Recent results show that photonics allows analog optical computing \cite{wu2021analog} or more general optical computing \cite{kazanskiy2022optical} and neuromorphic computing \cite{el2022photonic}. Remarkably, example of the use of photonic devices in edge-computing deep-learning applications appeared \cite{sludds2022delocalized}. Photonics Integrated Circuits (PICs) offer a potential solution to the limitations of electronic ANNs, as they enable high-speed, parallel transmission with low power dissipation \cite{Liu2021research,Sui2020review}. PICs allow keeping the same architecture of successful neuromorphic hardware, with interconnected neurons that receive multiple inputs, which are weighted, combined, and processed via nonlinear activation functions before being passed on to other neurons (Fig. \ref{fig:opticalNeuron}). 
PICs make these operations easy to implement, making large matrix multiplication fast and energy-efficient \cite{cheng2021photonic, moss2022photonic}. This advantage has led to the development of photonic accelerators for electronic ANNs \cite{zhou2022photonic, al2022scaling}. In PIC, connections between neurons are established via waveguides in an on-chip optical switching network \cite{Sui2020review,testa2019integrated}, where the optical signal can be modified using tunable waveguide elements like phase shifters or Mach Zehnder interferometers \cite{vivien2016handbook}. Different photonic technologies have been experimented to implement ANN \cite{dabos2022neuromorphic}, with silicon photonics being the most promising one \cite{xu2022silicon}.   

The complexity of  operations possible with ANNs depends on several factors, including the network topology, which determines how processing units (neurons) are connected. PICs can be used for a variety of ANNs, from simple structures (such as the single perceptron shown in Fig. \ref{fig:opticalNeuron} right \cite{Rosenblatt1958perceptron}) to complex ones \cite{guo2021integrated}. In addition, programmable PICs are proposed as reconfigurable circuits to implement ANN \cite{bogaerts2020programmable}. The optimal structure depends on the specific task and the amount and format of data to be analyzed \cite{brunner2021competitive}. ANNs of the feed-forward type are suitable for low-latency and fast-reconfigurability problems \cite{pai2020parallel}, while recurrence is needed for time-dependent tasks, for instance, in high-complexity problems where long and short-term memories play a key role \cite{medsker1999recurrent}. Photonic reservoir computing is easily implemented in PICs \cite{antokik2020large}, where random fixed connections between nodes are established, with training only performed in the output layer. Finally, the readout strategy is another key element that provides direct access to the information elaborated by the network, with the choice of optical or electrical readout depending on the topology of the network and the specific requirements of the task \cite{ma2021comparing}.
Optical systems inherently involve complex numbers, making them well-suited for implementing complex-valued neural networks \cite{hirose1994application,lee2022complex,hirose2012generalization}.  This is because the propagation of light in waveguides and its nonlinear interaction with various media are best described in the complex domain, where both the phase and amplitude of the electric field associated with the optical signal must be taken into account. While each complex number can be represented by two real numbers, a complex-valued ANN is not equivalent to a real ANN with twice as many parameters \cite{hirose1994application,hirose2012generalization}. This is because the rotatory dynamics of complex numbers comes into play when performing complex multiplication, resulting in a reduction of degrees of freedom compared to the case of completely independent parameters. PICs have the advantage of easily manipulating complex numbers, which combined with carefully selected nonlinear nodes and an effective readout strategy, allows for simple hardware implementations of ANNs \cite{mancinelli2022perceptron}. These perform demanding tasks that would otherwise require significantly higher costs using traditional ANNs. A successful example of this type of photonic ANN shows chromatic dispersion equalization at 10 Gbps for a 100 km long optical fiber in tiny silicon chip with less than 100 mW power dissipation \cite{staffoli2023equalization}. As a comparison, a similar function realized with a digital signal processor (DSP) might consume up to 1 W.

The nonlinear activation function is a crucial component of the nodes in ANNs, as it determines the output of each node, and therefore plays a fundamental role in the learning process \cite{dabos2022neuromorphic,el2022photonic}. PICs offer a range of choices for implementing these nonlinear functions, such as using the square modulus via a photodector \cite{agrawal2002receivers} or the inherent non-linearity of the material\cite{borghi2017nonlinear} or implementing a Semiconductor Optical Amplifier (SOA) within the neuron \cite{hamerly2019large,shen2017deep}. SOAs behave linearly for low optical input power, but exhibit strong nonlinearity and reach saturation for higher power values  \cite{agrawal2002amplifiers}, making them suitable for acting as a nonlinear node \cite{shi2019deep}. In silicon photonics, most ANNs use tunable Mach-Zehnder interferometers (MZIs) as the basic
building block \cite{banerjee2022characterizing,wang2022chip,clements2016optimal}. However, this review focuses on nonlinear nodes based on microring resonators (MRRs) \cite{bogaerts2012silicon,pavesi2021thirty}. 

First, in section \ref{sec:intro}, we  briefly summarize the physics of MRR with a particular emphasis on their linear and nonlinear responses. Their stationary, time-resolved, linear, nonlinear and spectral responses are described with a set of equations that allows to design their transfer function. Relevant parameters are introduced and modeled, with a particular emphasis on the role of free carriers, temperature and surface wall roughness in the MRR physics. Then, in section \ref{sec:NeuralPhotonicMicroresonator}, the use of MRR as a neuron is introduced. A similarity between the functioning of biological neurons and the nonlinear response of MRR to different stimuli is presented. We discuss different implementations of MRRs within ANNs based on how the neurons/nodes are distributed: the spatial, the temporal, and the wavelength domains. A particular use of MRR in ANN is their use as weight bank. This is discussed in section \ref{sec:weight_bank}, where it is shown that their characteristic spectral response and the tuneability of their resonance allow imprinting different positive and negative weights on the optical signal. In section \ref{sec:Hybrid}, it is shown that further functions to silicon photonic ANN can be implemented when hybrid approaches are used. Here, we discuss the use of phase change materials to implement non-volatile memory elements in ANN. Finally, in section \ref{sec:Conclusions}, we draw the conclusion of this review by summarizing the advantages of using MRR in ANN and indicating few promising research directions.

\section{Basic on linear and nonlinear silicon microring resonators} \label{sec:intro}
In this section, we discuss the general properties of MRRs and we show their dynamic/stationary optical response in the linear and nonlinear regimes. This will help to further understand the potential and limitations of a silicon MRR used as an active node of an ANN.

\subsection{Stationary regime and spectral properties}\label{subsec:properties} 
\begin{figure*}[t!]
	\centering
	\includegraphics[width=1\textwidth]{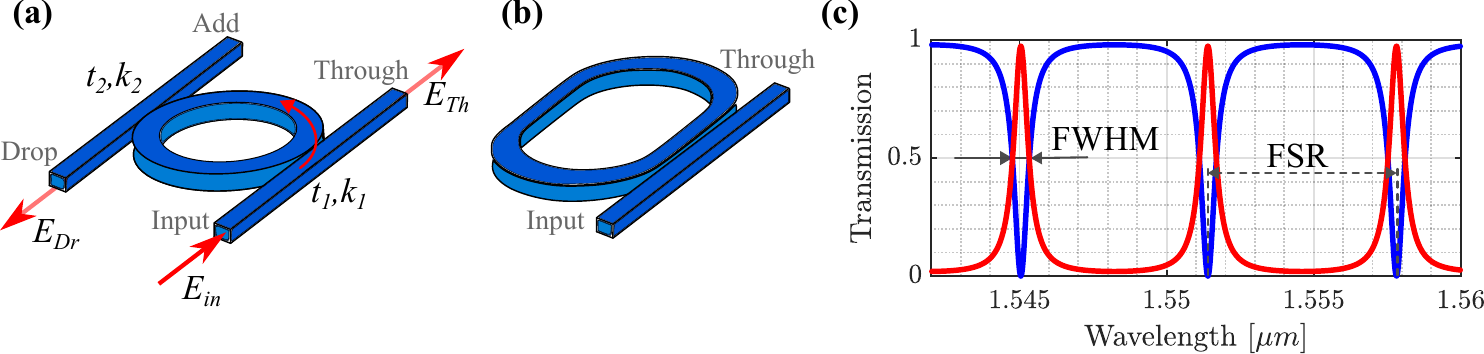}
	\caption{(a) and (b) sketch of a MRR in add-drop configuration and a race track MRR in all-pass one, respectively. Graph (c) shows the transmission spectra of the output ports of a MRR in the add-drop configuration. The blue line shows the output from the through port ($|E_{Th}/E_{in}|^2$), while the red one highlights the output from the drop port ($|E_{Dr}/E_{in}|^2$).}
	\label{fig:SketchRing}
\end{figure*}

The MRR has been extensively studied and described in literature \cite{heebner2008optical, bogaerts2012silicon, vahala2003optical, chrostowski2015silicon}. A MRR consists of a waveguide closed on itself to form a loop. In general, it can assume the form of any closed path. Typically, it resembles a circular ring or, if elongated with two straight sections in one direction, a racetrack circuit (see Fig. \ref{fig:SketchRing} (a) and (b), respectively). Hereunder, we will focus on a MRR; however, the derivations and results apply to microresonator of any shape. 
 
A MRR placed alone inside a PIC represents an isolated system. Therefore, it is necessary to excite the MRR by coupling light into it. The most common way is to use a directional coupler. It is obtained by approaching a waveguide (also called a bus waveguide) to the external rim of the MRR. The whole MRR/bus waveguide system can take two configurations: add-drop (see Fig. \ref{fig:SketchRing} (a)) and all-pass (see Fig. \ref{fig:SketchRing} (b)). The former when the MRR is coupled to two waveguides, while the latter when it is coupled to a single waveguide. Note that the bus waveguide enables not only the study of the MRR properties but also its implementation into complex photonic structures.

Let us consider the MRR/bus waveguides system in add-drop configuration shown in Fig. \ref{fig:SketchRing} (a). It can be described in the steady state through the transfer matrix approach \cite{saleh2019fundamentals}. Assuming that the ring and bus waveguides are all single-mode, the signal used to excite the system is identified only by the amplitude of the guided mode. The coupling regions for the evanescent field are described by two lossless and reciprocal directional couplers. They are characterized by two real-values coefficients $k_i$ and $t_i$, with $i=1,2$, which describe the coupling to the MRR and the transmission to the output, respectively. Reciprocity and losslessness impose that $|k_i|^2+|t_i|^2 = 1$. Assuming negligible back reflections in the waveguides (this is not always the case in silicon MRRs, see subsection \ref{subsec:Back}), the transmittance from the output ports (through and drop) reduces to:
\begin{align}\label{eq:Et-Ed}
T_{Th} = \frac{E_{Th}}{E_{in}} =& \frac{t_1 - e^{-\alpha p} e^{\frac{2 i \pi n_{eff} p}{\lambda}} t_2}{1 -  e^{-\alpha p} e^{\frac{2 i \pi n_{eff} p}{\lambda}} t_1 t_2}, & 
T_{Dr} = \frac{E_{Dr}}{E_{in}} =& -\frac{e^{\frac{-\alpha p}{2}} e^{\frac{2 i \pi n_{eff} p}{2 \lambda}}k_1 k_2}{1 -  e^{-\alpha p} e^{\frac{2 i \pi n_{eff} p}{\lambda}} t_1 t_2},
\end{align}
where  $E_{in}$, $E_{Th}$ and $E_{Dr}$ are the input field, the transmitted field at the through port and the output field at the drop port, respectively. $p$ is the perimeter of the ring circumference, $\alpha$ is the loss coefficient, $n_{eff}$ is the effective refractive index and $\lambda$ is the wavelength of the incident laser. Noteworthy, $T_{Th}$ can be formulated in the all-pass configuration by simply imposing $t_2=1$.

Figure \ref{fig:SketchRing}(c) shows the transmission spectra of an add-drop MRR, that is, the modulus square of Eq. \ref{eq:Et-Ed} as a function of wavelength. It exhibits a set of symmetrical Lorentzians having a peak in the drop port (red line) and a dip in the through one (blue curve). The maxima/minima of these Lorentzians are the resonance condition, namely, when the wavelength of the incident light fits an integer number of times inside the optical length of the MRR ($\lambda_0 = n_{eff} p / m$, with $m = 1,2,3, ...$). This condition ensures that the phase of the waves after a round trip is equal to a multiple of $2 \pi$, and consequently, it causes the waves to constructively interfere within the cavity. The distance between two resonances is defined as the free spectral range (FSR). Within a first-order approximation of the dispersion it is the wavelength range equal to \cite{bogaerts2012silicon}: 

\begin{equation}
    FSR=\frac{\lambda^2}{n_{g} p} ,
\end{equation} 
where $n_g$ is the group index. The FSR thus depends on the geometrical size of the system.

An outstanding spectral feature related to losses and coupling coefficients is the full width at half maximum (FWHM) of the Lorentzian. It can be formulated as \cite{bogaerts2012silicon}: 
\begin{equation}\label{eq:FWHM}
	FWHM = \frac{\left( 1- t_1 t_2 e^{-\frac{\alpha p}{2}}\right) \lambda_0^2}{\pi n_g p \sqrt{t_1 t_2 e^{-\frac{\alpha p}{2}}}},
\end{equation}
where $\lambda_0$ is the resonance wavelength. 
Note that both FSR and FWHM depend on the group index and not on the effective one. Since $n_g = n_{eff} - \lambda_0 \frac{dn_{eff}}{d \lambda}$, they take into account the waveguide dispersion and are therefore wavelength-dependent.  
The FWHM is closely related to the energy stored in the cavity and allows the spectral definition of the fingerprint of the MRR, i.e., the quality factor Q \cite{heebner2008optical}:
\begin{equation}\label{eq:Q}
	Q = \frac{\lambda_0}{FWHM}.
\end{equation}
Q describes the efficiency with which a MRR traps light. Advances in fabrication techniques have led to the realization of increasingly high-performance MRRs \cite{chrostowski2015silicon}, showing record values of Q close to one billion \cite{puckett2021422}.

\subsection{Time response and coupling regimes} \label{subsec:TCMT}
The physics of a MRR/bus waveguides system can be simply described by using the Temporal Coupled Mode Theory (TCMT) with the common time-reversal invariant system properties \cite{fan2003temporal, suh2004temporal}. This approach treats the system as spatially dimensionless and it is a powerful modeling tool that can be extended to more complex structures in both the linear regime and, as we will see in subsection \ref{subsec:NonLinear}, the nonlinear one. 
In the add-drop configuration, the temporal dynamics of a MRR mode with a field amplitude $a$, excited by an incident field whose amplitude is $E_{in}$, can be formulated as:
\begin{align}
	i \frac{da}{dt} =& \left(\omega_0 - i \left(\tilde{\gamma} + \Gamma_1\right)\right) a[t] - \sqrt{2 \Gamma_1} E_{in}[t], \\
	E_{Th}[t] =& E_{in}[t] + i \sqrt{2 \Gamma_1} a[t], \label{eq:TCMT}\\
	E_{Dr}[t] =& e^{i \pi m} i \sqrt{2 \Gamma_2} a[t].  	
\end{align}
$\omega_0 =\frac{2 \pi c}{\lambda_0}$ is the resonant angular frequency of the MRR, c the speed of light in vacuum and $\tilde{\gamma} = \gamma + \Gamma_2$. $\Gamma_1$ and $\Gamma_2$ are the extrinsic damping rates, which are related to the coupling with the top and bottom bus waveguides, respectively. $\gamma$ is the intrinsic damping rate and it describes the losses in the MRR such as material absorption, bending and scattering. The exponential term in $E_{Dr}$ takes into account the different relative phase for the even and odd modes of the MRR.    

\begin{figure*}[t!]
	\centering
	\includegraphics[width=1\textwidth]{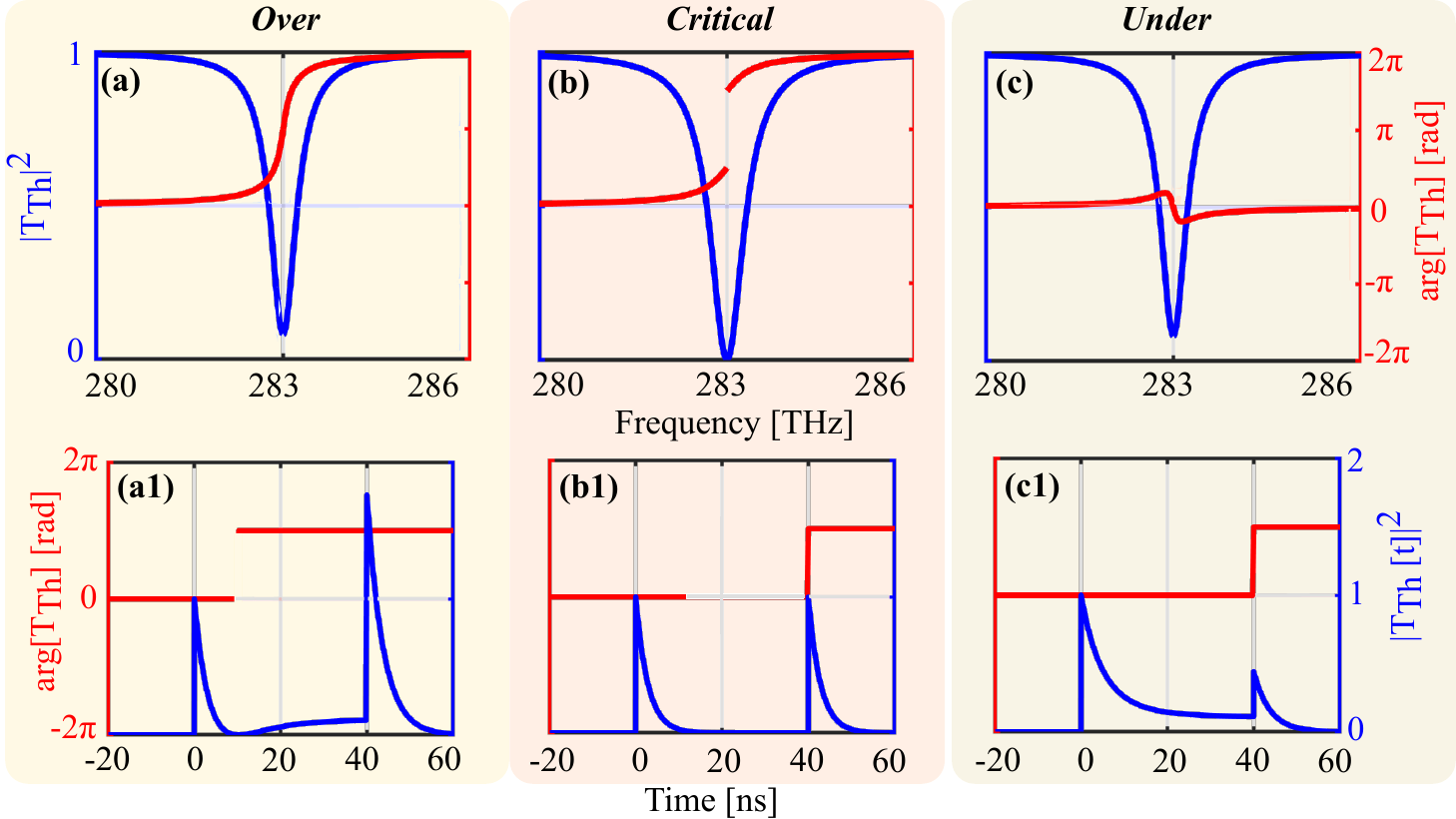}
	\caption{(a)-(a1), (b)-(b1) and (c)-(c1) transmission (blue line) and phase (red line) as a function of time and frequency for the over-coupling, critical-coupling and under-coupling regime, respectively. The temporal profile shows the response of the MRR/bus waveguide system in resonance ($\Delta \omega = 0$) observed at the through port for a rectangular input pulse which lasts from 0 to 40 ns.}
	\label{fig:StaTime}
\end{figure*}
Considering a monochromatic input field ($E_{in} = \epsilon_{in} e^{-i \omega t}$) and a steady state solution of the type $a = \tilde{a} e^{-i \omega t}$, the previous system of equations leads to the following steady-state transmittance at the output ports:
\begin{align}\label{eq:TTh-TdTime}
	T_{Th} = \frac{E_{Th}}{E_{in}} =& 1 - \frac{2 \Gamma_1}{-i \Delta \omega + \tilde{\gamma} + \Gamma_1}, & 
	T_{Dr} = \frac{E_{Dr}}{E_{in}} =& - \frac{e^{i \pi m} 2 \sqrt{\Gamma_1\Gamma_2}}{-i \Delta \omega + \tilde{\gamma} + \Gamma_1},
\end{align}
where $\Delta \omega = \omega - \omega_0$. In accordance with the transmission of subsection \ref{subsec:properties}, the modulus square of these equations resembles a Lorentzian shape: 
\begin{align}\label{eq:IntensityTime}
	|T_{Th}|^2 =& 1 - \frac{4 \tilde{\gamma} \Gamma_1}{\Delta \omega^2 + \left(\tilde{\gamma} + \Gamma_1\right)^2}, & 
	|T_{Dr}|^2 =& \frac{4 \Gamma_1 \Gamma_2}{\Delta \omega^2 + \left(\tilde{\gamma} + \Gamma_1\right)^2}.
\end{align} 
One can also obtain Eq. \ref{eq:IntensityTime} as an expansion around a single resonance of the transmittances formulated by the transfer matrix approach (Eq. \ref{eq:Et-Ed}). As a result, $\gamma =\frac{1-e^{-2\alpha p}}{2 n_g p } c$ and $\Gamma_{i} = \frac{k_i^2 c}{2 n_g p}$ with $i=1,2$.

Panels (a), (b) and (c) in Fig. \ref{fig:StaTime} show $|T_{Th}|^2$ and its phase ($arg[T_ {Th}]$) as a function of the wavelength for different values of the intrinsic and extrinsic coefficients, corresponding to the regime of over-coupling ($\Gamma > \tilde{\gamma}$), critical-coupling ($\Gamma = \tilde{\gamma}$) and under-coupling ($\tilde{\gamma} < \Gamma$).  Since Eqs. \ref{eq:IntensityTime} are symmetric for an exchange of $\tilde{\gamma}$ with $\Gamma$ and \textit{vice versa}, the transmitted intensity does not allow distinguishing between the under- and over-coupling regime (see blue curve of panels (a) and (c)). In contrast, $\arg[T_{Th}]$ exhibits a different behavior by swapping $\Gamma$ with $\tilde{\gamma}$. Specifically, in resonance the phase is zero for the under-coupling regime while it is $\pi$ for the over-coupling one (see red curves of panels (a) and (c)). The reason for this difference is related to the interaction between the propagating field in the bus waveguide and that exiting the MRR (Eq. \ref{eq:TCMT}). It can be easily understood by looking at the time-dependent transmission of the MRR/bus waveguide system. In the linear regime, the time evolution can be obtained through the Green function \cite{duffy2015green}. As a result, the field at the output ports reduces to the convolution between the Green function of the MRR and the input field \cite{biasi2019time}. 

Panels (a1), (b1) and (c1) of Fig. \ref{fig:StaTime} show the transmission (blue curves) and the phase (red lines) as a function of time at resonance for a rectangular pulse excitation, in the over-coupling, critical-coupling and under-coupling regimes, respectively. The temporal dynamics allow to distinguish between the different regimes only by looking at the transmission. It also shows how the destructive interference reduces the field to zero in the critical-coupling regime and determines the value of the phase of the output field. The charging and discharging times are featured by an exponential drop. Its time constant is directly related to the energy stored in the MRR and permits to define \cite{biasi2019time}:
\begin{equation}\label{eq:Qtime}
	Q= \frac{\omega_0}{2\left( \tilde{\gamma}+\Gamma_1 \right)}.
\end{equation}
Noteworthy, $Q$ is directly proportional to the lifetime of the photon in the MRR, defined as $\tau = 1/ \left( \tilde{\gamma} +\Gamma_1 \right)$. The longer the photon lifetime, the greater the enhancement of the stored optical power. As a result, the experimental estimation of Q through cavity ring down techniques \cite{berden2009cavity} becomes possible with standard setups.

A further relevant parameter describing the performance of a MRR/bus waveguides system is the power enhancement factor, F$_e$. It is defined as the ratio between the power circulating in the MRR and the incoming power \cite{borghi2017nonlinear}: 
\begin{equation}\label{eq:Fe}
	F_e = \frac{2 c \, \Gamma_1}{p \, n_g \left( \tilde{\gamma} + \Gamma_1 \right)^2} = \frac{8 c \Gamma_1}{p n_g} \left(\frac{Q}{\omega_0}\right)^2.
\end{equation}
$F_e$ can take values ranging from a few tens up to tens of thousands, so that watts of power can be circulated in the MRR with sub-milliwatt excitation. Consequently, high-Q MRRs are ideal platforms to induce strong light-matter interaction, which dramatically enhances nonlinear phenomena.  


\subsection{Nonlinear regime} \label{subsec:NonLinear}
The MRRs exhibit a strong increase in steady power density close to the resonance conditions. As a result, also for low overall optical power, the polarization vector ($\mathbf{P}$) can no longer be approximated as directly proportional to the applied optical field ($\mathbf{E}$), but has to be formulated as \cite{boyd2020nonlinear}:
\begin{equation}\label{eq:P}
	\mathbf{P} = \epsilon_0 \left(\chi^{(1)} \cdot \mathbf{E} + \chi^{(2)} \cdot \mathbf{E}^2+ \chi^{(3)} \cdot \mathbf{E}^3 + ... \right),
\end{equation}
where $\epsilon_0$ is the vacuum permittivity, $\chi^{(i)}$ is the i-th susceptibility tensor. Since silicon has a centrosymmetric structure, it does not show native second-order nonlinearities ($\chi^{(2)}=0$) \cite{boyd2020nonlinear,leuthold2010nonlinear} unless it is engineered \cite{castellan2019origin,franchi2020second}. Consequently, we will focus on the nonlinear effects related to $\chi^{(3)}$. Indeed, the non-linear terms given by higher-order susceptibilities are smaller and smaller and only effective for very large field intensities \cite{borghi2017nonlinear}.

A relevant outcome of third-order nonlinear processes is an intensity-dependent refractive index \cite{leuthold2010nonlinear}:
\begin{equation}\label{eq:n_nonl}
	n = n_0 + n_2 I - i \frac{\lambda}{4 \pi} \left(\alpha_0 + \alpha_2 I \right) ,
\end{equation}
where $I$ is the field intensity, $n_0$ and $\alpha_0$ are the linear wavelength-dependent refractive index and absorption, respectively. $n_2$ is the Kerr coefficient and $\alpha_{2}$ is the Two Photon Absorption (TPA) one. These are respectively linked to the real and imaginary part of $\chi^{(3)}$ as:
\begin{align}\label{eq:KerrAndTPA}
	n_2=& \frac{3}{4 c n_0^2 \epsilon_0} \Re e \left[\chi^{(3)} \right], & 
	\alpha_2 =& \frac{- 3 \omega}{2 c^2 n_0^2 \epsilon_0} \Im m \left[\chi^{(3)} \right].
\end{align}
TPA is due to the absorption of two photons, whose energy sum matches the excitation of an electron from the valence band to the conduction one \cite{boyd2020nonlinear}. Therefore, TPA generates free carriers in the conduction and valence bands that consequently cause Free-Carrier Absorption (FCA) and Free-Carrier Dispersion (FCD) processes. These are related to the first-order susceptibility. FCA and FCD introduce a change in both the real and imaginary parts of the refractive index \cite{borghi2017nonlinear}. More precisely, the FCA process changes the imaginary part, while the FCD process modifies the real part.  The absorption of light and the thermalization of free carriers induce heat in the waveguides that gives rise to an increase of the silicon temperature which reflects in a change of the refractive index due to the thermo-optic effect. 

The nonlinearity of silicon and the related processes enrich the dynamics of a MRR/bus waveguide system. The temporal Eq. \ref{eq:TCMT_2} has to be updated by a dimensionless resonance shift ($\delta$) that accounts for FCD, Kerr nonlinearity and thermo-optic effect. In addition to the time evolution of the field amplitude ($a$), the temporal dynamics of the free carrier population ($\Delta N$) and that of the temperature difference between the waveguide core and the cladding ($\Delta T$) have to be considered \cite{johnson2006self, pernice2010time}:
\begin{align}\label{eq:Field}
    i \frac{{\rm d}a}{{\rm d}t} = 
    &\left[\omega_0\left(1+\delta\right) - i \left(\tilde{\gamma} +\Gamma_1\right) \right] a - \sqrt{2 \Gamma_{1}} E_{in}\,,\\
    \label{eq:FreeCarrier}
    \frac{{\rm d}\Delta N}{{\rm d}t} =
    &-\frac{\Delta N}{\tau_{fc}}+g_{\rm TPA} \left| a \right|^4,\\
    \label{eq:Temprature}
    \frac{{\rm d} \Delta T}{{\rm d}t} =
    &-\frac{\Delta T}{\tau_{th}} + \frac{P_{\rm abs}}{m\,c_p}\,,\\ 
    \label{eq:delta}
    \delta = 
    &\!-\!\frac{1}{n_0}\!\!\left[\!\frac{{\rm d} n}{{\rm d} T} \Delta T \!+\! \sigma_{\rm FCD} \Delta N \! + \! n_2 \left|a\right|^2\!\right]\!\!,\\
    \label{eq:loss}
    \gamma = 
    &\Gamma_1+\tilde{\gamma}+\eta_{\rm FCA}\Delta N + \eta_{\rm TPA} \left|a \right|^2.
\end{align}
$g_{\rm TPA}$ and $\eta_{\rm TPA}$ are the free carrier generation rate and the TPA loss rate per unit energy, respectively. $\eta_{\rm FCA}$ is the loss rate due to free carrier absorption. These coefficients are defined in \cite{zhang2013multibistability, borghi2017nonlinear}. $m$  is the MRR mass and $c_p$ the specific heat of silicon. $\sigma_{FCD}$ is a negative coefficient denoting the free carrier dispersion and $dn/dT$ denotes the positive thermo-optic coefficient of silicon. Along with the quasi-instantaneous Kerr effect (ruled by $n_2$) they determine $\delta$ (equation \ref{eq:delta}) by generating two contrasting shifts of the resonant frequency. Noteworthy, these are characterized by two different relaxation times: the thermal relaxation time $\tau_{fc}$ and the free carrier lifetime $\tau_{th}$. Furthermore, they also show a distinct field amplitude dependence. Indeed, $\Delta N$ is governed by the square of the field intensity. Differently, $\Delta T$ scales with the absorbed power and, since $P_{abs} = 2 (\gamma - \Gamma_1 - \Gamma_2) |a|^2$, it is directly proportional to the field intensity. Typically, the thermal and carrier lifetimes are respectively of the order of $\tau_{th} \simeq (60-150) \,ns$ and $\tau_{fc} \simeq (1-10) \,ns$ \cite{van2012simplified}. However, other fabrication processes can yield different values as demonstrated in a recent work \cite{borghi2021modeling}, where $\tau_{th} \simeq 280 \, ns$ and $\tau_{fc} \simeq 45 \, ns$.

\begin{figure*}[t!]
	\centering
	\includegraphics[width=1\textwidth]{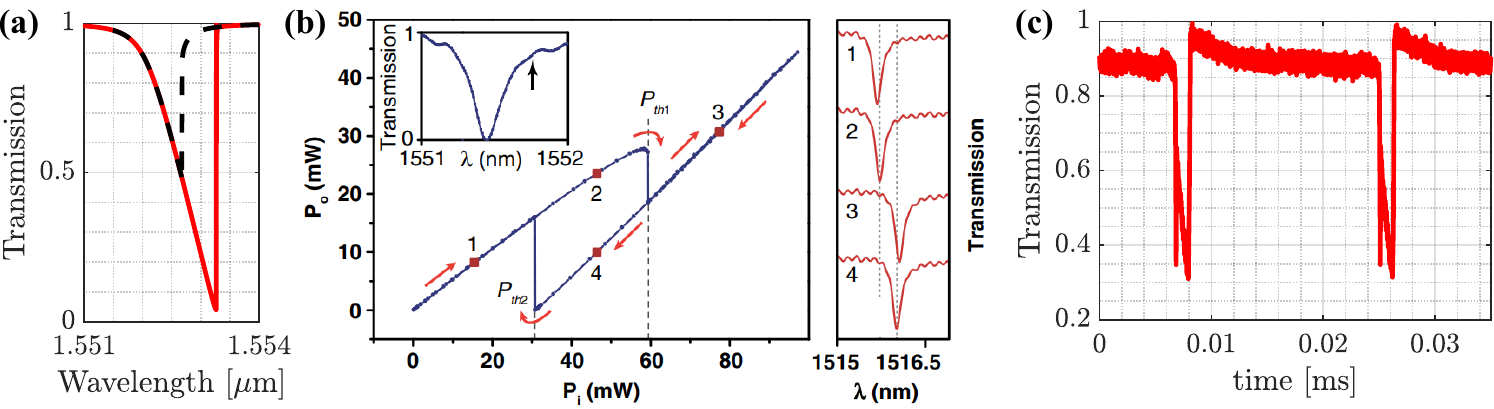}
	\caption{(a) Transmission as a function of the input wavelength for wavelength sweep in upwards (red lines) and downwards (black-dashed lines) directions. (b) Transmitted power as a function of input power. The inset displays the cold resonance spectrum. Here, the black arrow denotes the wavelength of the input laser. The red arrows indicate the increasing and decreasing of the input power around the points: 1,2,3 and 4. On the right, it is reported the stable transmitted state for each power corresponding to these points. Reprint by permission from \cite{ramiro2013thermo}. (c) Through port transmission for a silicon MRR in the nonlinear regime when the self-pulsing oscillation is established.}
	\label{fig:NonLinResp}
\end{figure*} 
The transmission spectrum of the MRR/bus waveguide system is no longer a simple Lorentzian at high powers. The silicon nonlinearity causes a resonance shift related to the internal power inside the MRR. As a result, the line shape changes and assumes a typical triangular shape (see Fig. \ref{fig:NonLinResp} (a)). The jump shown immediately after the resonance is a typical feature of optical bistability \cite{boyd2020nonlinear, de2021nonlinearity, priem2005optical}. The bistable loop can be observed by comparing the transmission spectrum along a wavelength ramp in the upward and downward directions at sufficiently high power, as shown in Fig.  \ref{fig:NonLinResp} (a). Here, the MRR exhibits a distinct response for the two directions of the wavelength sweep. Similarly, as shown in Fig. \ref{fig:NonLinResp} (b), a hysteresis loop in the power spectrum is observed by fixing a wavelength in the bistable region \cite{almeida2004optical}. Such bistability can be based on either the thermal nonlinear optical effect \cite{almeida2004optical} or the FCD effect generated by the TPA \cite{xu2006carrier}. 
The two main differences between these two nonlinearities are related to the different direction of the Lorenzian shift (toward red for thermal and toward blue for FCD) and the different time scale. The free carrier effect is much faster than the thermo-optic one.

The combination of the thermo-optic, TPA and FCD effects can result in self-pulsing behavior \cite{johnson2006self}. In this case, an input CW signal with a wavelength close to the resonant wavelength of the MRR is converted into a periodic oscillating output signal. An example of this experimental response for the through port is shown in Fig. \ref{fig:NonLinResp} (c). The first sharp dip is related to the free carrier dynamics while the second broad one is due to the nonlinear thermal effect. While in a single MRR the dynamics is well described and generates a deterministic effect \cite{borghi2021modeling}, in a system composed of several cascaded MRRs (as in a SCISSOR, side-coupled integrated spaced sequences of resonators) it can assume a chaotic behavior \cite{mancinelli2014chaotic}. 

\begin{figure*}[t!]
	\centering
	\includegraphics[width=1\textwidth]{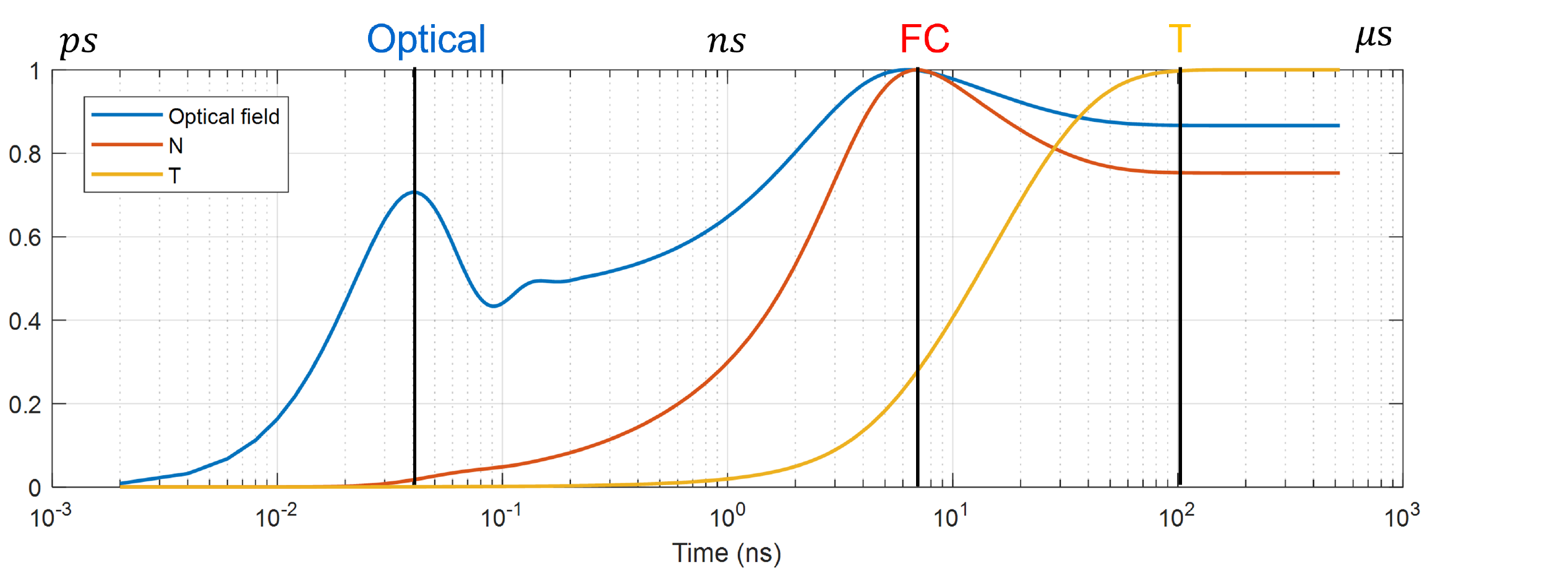}
	\caption{Optical field intensity at the drop port (blue line), variation of the free carrier concentration (orange line), and the temperature (yellow line) in normalized units as a function of time. The evolution of the intensity results from three different temporal dynamics, i.e. the optical charging of the MRR, the free carriers generation by TPA, and the temperature variation due to free carrier relaxations. The different curves were obtained numerically by solving the system of nonlinear Eq. \ref{eq:Field}-\ref{eq:loss} with a CW input, a detuning of 10 GHz and a coupling coefficient to the MRR of $k^2$=2 \%.
 }
	\label{fig:NonLinearTemporalEvolution}
\end{figure*}

It is worth emphasizing that a MRR allows implementing dynamical responses that have different time scales. Figure \ref{fig:NonLinearTemporalEvolution} shows the time response of a MRR in the add-drop configuration, obtained with the system of nonlinear Eq. \ref{eq:Field}-\ref{eq:loss}. In the simulations, the input field is constant with an input frequency set to $\Delta \omega = 2 \pi\cdot 10\, \rm GHz$. The output field intensity at the drop port shows a rapid increase in a time scale of few hundreds of picoseconds due to the MRR charging (see section \ref{subsec:TCMT}). At this time, the change in temperature and free carrier concentration is negligible. Thereafter, the temporal evolution of the optical field is mainly related to the increase of the free carrier concentration due to TPA in the MRR. The free carrier concentration increase occurs in a time of few ns. This leads to a change in temperature, which further affects the temporal dynamics of the optical intensity, mainly from 10 ns to 100 ns. Finally, the optical intensity, as well as the free carrier concentration and temperature, reaches a steady state value at about 100 ns. As we will discuss in section \ref{sec:NeuralPhotonicMicroresonator}, the different temporal dynamics can be used in ANNs both to impress a fading memory and to obtain specific nonlinear responses of the MRR as an active network node.

\subsection{The backscattering of the surface-wall roughness} \label{subsec:Back}
\begin{figure*}[t!]
	\centering
	\includegraphics[width=1\textwidth]{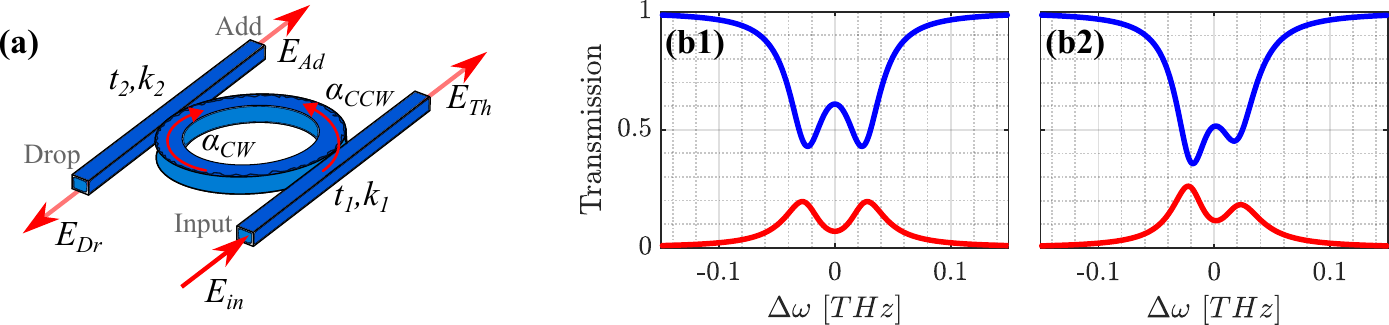}
	\caption{(a) Sketch of a MRR in add-drop configuration. The arrows indicate the propagating fields and the different symbols are defined in the text. Panels (b1) and (b2) show the outgoing intensity from the through and drop port as a function of detuning for a Hermitian and non-Hermitian coupling, respectively.}
	\label{fig:SketchBack}
\end{figure*}

In the ideal case, injecting light from one edge of the bus waveguide excites a mode in the MRR with a well-defined propagation direction. Consequently, in the add-drop configuration, there is no outgoing field from the input port ($E_{r} = 0$, where $E_{r}$ is the reflected field at the input port) and the add one ($E_{Ad} = 0$). However, in the non-ideal case, the fabrication process causes roughness on the waveguide surfaces, which induces backscattering of light and, therefore, excitation of counter-propagating modes \cite{gorodetsky2000rayleigh, li2016backscattering}. When the fabrication process is the limiting source of losses (typically in high or ultra-high Q MRRs), the steady state of the system is no longer a Lorentzian but a resonant doublet \cite{biasi2018hermitian, mccutcheon2021backscattering} (see Fig. \ref{fig:SketchBack} (b)). Thus, the repeated back-scattered light process induces a steady-state super-mode composed of counter-propagating modes, which exhibits a significant splitting in the transmission spectra.

An analytical model capable of catching the physics of surface-wall roughness backscattering in a MRR/bus waveguides system can be based on the TCMT seen in subsections \ref{subsec:TCMT} and \ref{subsec:NonLinear}. The coupling between the amplitudes of the clockwise ($\alpha_{CW}$) and counterclockwise ($\alpha_{CCW}$) propagating modes is defined by the complex coefficients $\beta_{12}$ and $\beta_{21}$. As a result, the temporal dynamics reduces to \cite{biasi2018hermitian}:    
\begin{equation}\label{eq:TCMT_2}
		i \frac{d}{dt} \begin{pmatrix}
			\alpha_{CCW} \\
			\alpha_{CW}
		\end{pmatrix}
		= 
	 \begin{pmatrix}
			\omega_0-i(\tilde{\gamma}+\Gamma_1) & -i\beta_{12} \\
			-i\beta_{21} & \omega_0-i(\tilde{\gamma}+\Gamma_1)
		\end{pmatrix}
		\begin{pmatrix}
			\alpha_{CCW} \\
			\alpha_{CW}
		\end{pmatrix}
		- \sqrt{2\Gamma_1}
		\begin{pmatrix}
			E_{\rm in} \\
			0
		\end{pmatrix}.
\end{equation}
The output fields highlighted in Fig. \ref{fig:SketchBack} (a) are defined as: $E_{Th}[t] = E_{in}[t] + i \sqrt{2 \Gamma_1} \alpha_{CCW}[t] $, $E_{Dr}[t] = e^{i \pi m} i \sqrt{2 \Gamma_2} \alpha_{CCW}[t]$, $E_{Ad}[t] = e^{i \pi m} i \sqrt{2 \Gamma_2} \alpha_{CW}[t]$ and $E_{r}[t] = i \sqrt{2 \Gamma_1} \alpha_{CW}[t]$. The off-diagonal coupling coefficients can take distinct values, inducing a reflected field at the input. In contrast, the Lorentz reciprocity theorem ensures that exciting the left or right edge of the waveguide does not change the transmission response. To schematize the huge number of values that the coupling coefficients can assume, the inter-modal Hermitian ($h = i (\beta_{12}-\beta_{21}^*)/2$ ) and non-Hermitian ($n = i (\beta_{12}+\beta_{21}^*)/2$ ) coefficients are defined. In order to avoid gain in the MRR/bus waveguides system, $|n| \leq \tilde{\gamma}$ \cite{biasi2018hermitian}.

A monochromatic input field ($E_{in} = \epsilon_{in} e^{-i \omega t}$) leads to the following stationary solutions for the transmittance of the output ports: 
\begin{align}\label{eq:Et-EdTime}
	T_{Th} = \frac{\epsilon_{Th}}{\epsilon_{in}} =& 1 - \frac{2 \Gamma_1 \left(-i \Delta \omega + \tilde{\gamma}+ \Gamma_1\right)}{\left(-i \Delta \omega + \tilde{\gamma} + \Gamma\right)^2-\beta_{12} \beta_{21}}, &
	T_{Dr} = - \frac{e^{i \pi m}2 \sqrt{\Gamma_1 \Gamma_2} \left(-i \Delta \omega + \tilde{\gamma}+ \Gamma_1\right)}{\left(-i \Delta \omega + \tilde{\gamma} + \Gamma\right)^2-\beta_{12} \beta_{21}}.
\end{align} 
Figure \ref{fig:SketchBack} (b1) and (b2) show the doublet of the transmission spectrum from the through and drop ports (blue and red curves, respectively) for Hermitian and non-Hermitian couplings. In the Hermitian (conservative) case, $\beta_{12} = -\beta_{21}^*$ holds, and the stationary response shows a symmetric doublet, namely, with the same valley depths and peak heights. This is the result of the same energy exchange between the counter-propagating modes. In the non-Hermitian (non-conservative) case, there is no relation between $\beta_{12}$ and $\beta_{21}$. Consequently, the asymmetric exchange of energy between the clockwise and counterclockwise leads to an unbalanced doublet (Fig. \ref{fig:SketchBack} (b2)).

The interaction between the counter-propagating modes modifies the time evolution of an ideal MRR shown in Fig. \ref{fig:StaTime}. It gives rise to fast oscillations over the exponential decays in the charging and discharging phase, making the classical estimation of Q prohibitive \cite{biasi2022interferometric}. The nonlinear dynamics of a MRR affected by backscattering is complex to model and requires the introduction of the counter-propagating mode equation into the TCMT, i.e., Eq. \ref{eq:Field},\ref{eq:FreeCarrier} and \ref{eq:Temprature}.

\section{Neural networks based on silicon photonics microresonators}
\label{sec:NeuralPhotonicMicroresonator}
This section discusses the use of silicon MRRs in photonic ANNs. The parallelism between the nonlinear response of a MRR and the behavior of a biological neuron is explored, and then the implementation of MRR within ANNs is considered. We review different approaches ranging from a MRR as a spiking neuron to a MRR as a source of frequency combs. Typically, at each layer of an ANN the input data is mapped into a higher-dimensional space and processed by nonlinear functions. The mapping can be viewed as a linear matrix-vector multiplication with certain weights.
The size of the input domain can be increased i) spatially by distinct physical topologies, ii) temporally using virtual nodes and iii) in wavelength via wavelength multiplexing techniques. These methods can be combined in complex hyperspaces increasing the performance of the network. The nonlinear response of a MRR can be used to nonlinearly transform the input signal or to generate a fading memory in the network. Typically, the distinction between nonlinearity and induced fading memory is not well defined and strongly depends on the optically encoded information. In the following, we show how the nonlinear MRR response seen in section \ref{subsec:NonLinear} has been used in the ANN through the leveraging of these distinct domains. 

\subsection{Nonlinear response of a microresonator as an activation function}
\label{subsec:spike}
Biological neurons exhibit an enormous diversity of spiking activities and can be thought of as dynamic nonlinear systems that respond to an impulse through their intrinsic behavior \cite{izhikevich2007dynamical}. As a result, a qualitative description of their dynamics can be obtained by studying the phase portraits (geometrical representations of the dynamical trajectories followed by systems in the phase space) \cite{strogatz2018nonlinear}. In particular, the phase portraits show certain special trajectories that determine the topological behavior of all other trajectories in the phase space. For a given set of initial conditions, the phase portrait asserts whether an attractor, repellor and/or limit cycle occurs in the chosen path. An attractor is a stable point of the system, a repellor is an unstable one and the limit cycle is an isolated periodic orbit.

From the perspective of nonlinear dynamical systems, neurons become excitable when they are near a transition from a resting state to a state characterized by spiking activity. Such a transition is called a bifurcation. Surprisingly, there can be a huge number of possible mechanisms of excitability and spiking, but there are only four different types of equilibrium bifurcations that such systems can follow: \textit{saddle-node on invariant circle}, \textit{saddle-node}, \textit{subcritical Andronov-Hopf} and \textit{supercritical Andronov-Hopf} \cite{izhikevich2007dynamical}. Only the saddle-node and subcritical Andronov-Hopf bifurcations show a coexistence of a resting and a spiking state. It is revealed by a bistability, i.e., hysteresis behavior when increasing and decreasing input is injected; in fact, transitions from resting to spiking or from spiking to resting occur at different current values \cite{izhikevich2007dynamical}. The type of bifurcation defines the major neurocomputational properties \cite{rinzel1998analysis}.

The saddle-node on invariant circle and saddle-node bifurcations are features of neurons called \textit{integrators}, while the subcritical Andronov-Hopf and supercritical Andronov-Hopf are features of neurons called \textit{resonators} \cite{izhikevich2007dynamical}. Integrators and resonators represent two main modes of neuron activity. The integrators have a well-defined firing threshold and all-or-none spiking dynamics. As a result, if the perturbation is above a defined threshold spikes are observed, otherwise not. These spikes are generally stereotypical with a relatively constant amplitude. In contrast, resonators are characterized by subthreshold amplitude oscillations and do not show a sharp threshold but instead a smooth threshold region. They generate spikes with an amplitude that depends on the strength of the perturbation only for a certain range of powers. More importantly, resonators can exhibit spikes even with weak perturbations that are, nevertheless, resonant with the subthreshold oscillation frequency. This occurs regardless of whether the stimulation is excitatory or inhibitory.

According to Hodking's classification, resonators belong to \textit{class II} excitability \cite{fitzhugh1955mathematical}. Indeed, the responses or action potentials of the class II’s neurons are generated in a certain frequency band that is relatively insensitive to the variation of the applied current strength \cite{izhikevich2007dynamical}. Differently, integrators can also belong to the \textit{class I} if they correspond to a saddle-node bifurcation on invariant circle. The class I is defined by neurons that exhibit action potentials, which can be generated with an arbitrarily low frequency, depending on the applied current strength \cite{izhikevich2007dynamical}.

The dynamics of a passive MRR response can be studied in the language of nonlinear systems. To gain insight, the TCMT (equations \ref{eq:Field},\ref{eq:FreeCarrier} and \ref{eq:Temprature}) are used to construct 2D-phase portraits. Such phase map analysis allows to determine both the bifurcation and the excitability class of the MRR/bus waveguides system. The 2D-phase portraits are drawn by projecting the time traces of the propagating field amplitude ($a$), the temperature difference ($\Delta T$) and the free carrier population ($\Delta N$) in the $( \Delta T, N)$ plane for a certain power and wavelength of the input laser coupled top the MRR \cite{yacomotti2006fast}. In \cite{van2012cascadable}, it is shown that it is possible to reduce the dimensionality of the system by adiabatic elimination of the field variable $a$. Even neglecting the TPA contribution in the total loss ($\gamma$), Eq. \ref{eq:loss}, the relevant physics is preserved and the MRR still exhibits the self-pulsing oscillations. This reduction in the number of variables not only simplifies the study of phase portraits but also speeds up the numerical solutions. 

\begin{figure*}[t!]
	\centering
	\includegraphics[width=1\textwidth]{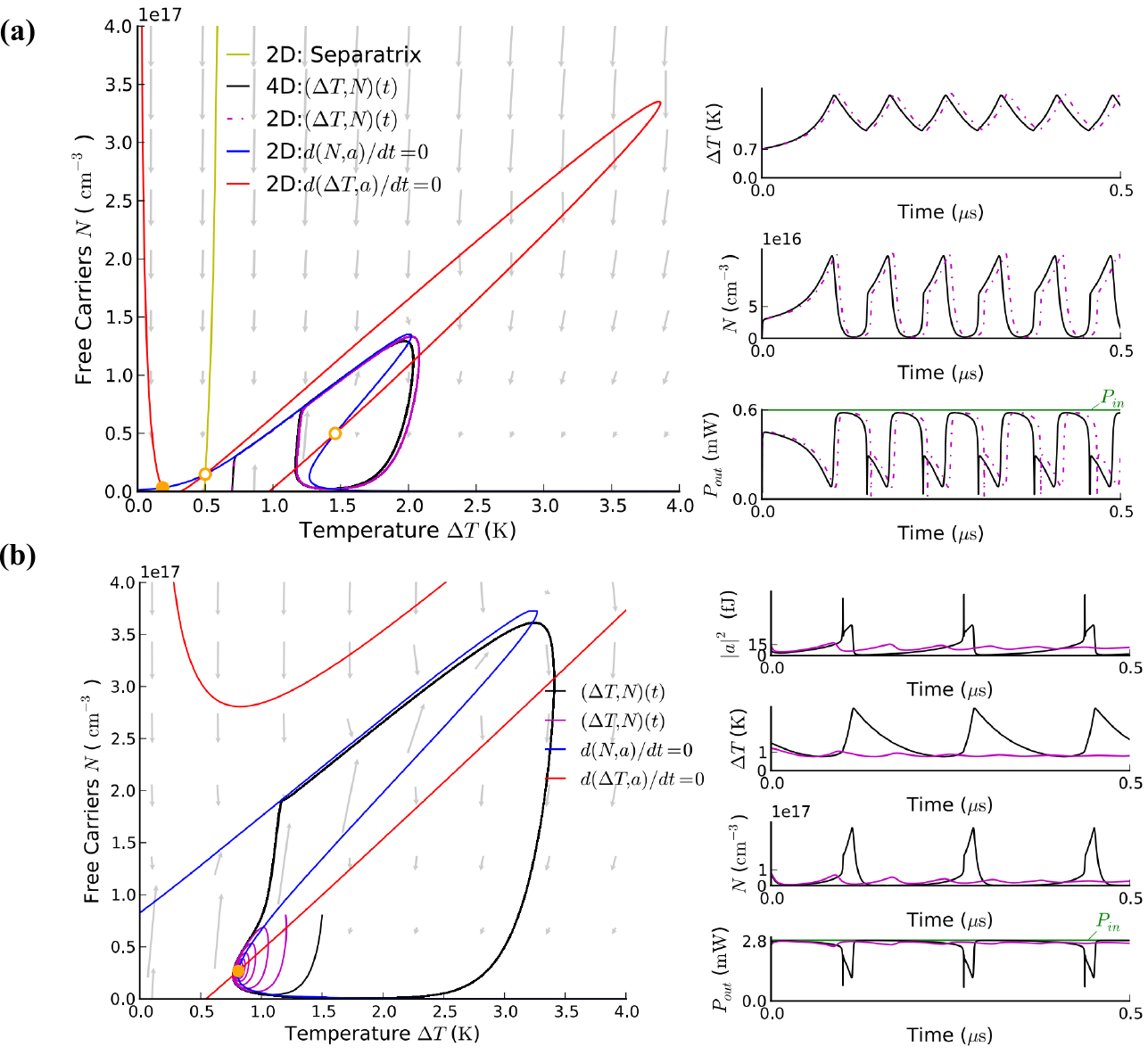}
	\caption{Phase portraits for a MRR (left) and corresponding time traces (right). (a) Comparison between the 4D model (black lines) and the approximated 2D model (dashed magenta lines). The phase portraits (left) show that the curves obtained with the complete model and the simplified one are comparable. This agreement is most evident in the time traces of the free carriers, temperature and optical power (right). The 2D model follows the trend of the 4D model, thereby capturing the physics of the system. (b) Results for a subcritical Andronov-Hopf bifurcation, where, depending on the initial conditions, one can obtain both the response highlighted by the black curves or the one denoted by the magenta lines. Reprint by permission from \cite{van2012cascadable}.}
	\label{fig:PhasePortrait}
\end{figure*}
Figures \ref{fig:PhasePortrait} (a) and (b) show the phase portraits for a power of 0.6 mW at a detuning $\Delta \lambda = \lambda -\lambda_{0} =$ 62 pm, and for a power of 2.85 mW and $\Delta \lambda = $-16 pm, respectively. Here, the fixed points are denoted by filled-orange circles and open-orange circles if they are stable and unstable, respectively. In (a) three fixed points are inferred of which one is stable and two are unstable. Both the curve of the approximated 2-D model (dashed magenta), i.e., neglecting the TPA term in the $\gamma$, and the curve of the non-approximated 4-D model (solid-black) are shown. The pattern is comparable in agreement with the corresponding time traces generated in the plots on the right. They show $\Delta T$, $N$, and $P_{out}$ (the power  output from the through port) as a function of time.  Again, the black line of the 4D model can be compared with the dashed magenta line of the 2D model. The phase portrait analysis states that for a given input power, the MRR can have one, two, or three fixed points. If the characteristics of the MRR lead to having two fixed points, it is subject to a saddle-node bifurcation. Conversely, if the MRR has two or three unstable fixed points, a stable limit cycle around the highest $|a|$ fixed point occurs.  However, if the MRR has three fixed points, at least one is stable (at low $|a|$).

In Fig. \ref{fig:PhasePortrait} (b), for a detuning toward the blue, there is a stable fixed point enclosed within a limit cycle. This configuration indicates the presence of a subcritical Andronov-Hopf bifurcation, which is thus characterized by a hysteresis cycle with respect to the input power.  The time traces on the right show that, as a function of the initial conditions $(\Delta T, N)(t=0)$, the trajectory converges to the limit circle showing the oscillating pattern in the black curve, or it either converges to the stable point showing the magenta response. Note that if the light is tuned toward red, in the bistability region, a different behavior can be obtained. This has no hysteresis cycle and was first classified as a supercritical Andronov-Hopf bifurcation in \cite{johnson2006self}. Here, it was shown that tuning the pump laser resonance from the blue side towards the resonance induces a supercritical Andronov-Hopf bifurcation that results in a limit cycle.

\begin{figure*}[t!]
	\centering
	\includegraphics[width=1\textwidth]{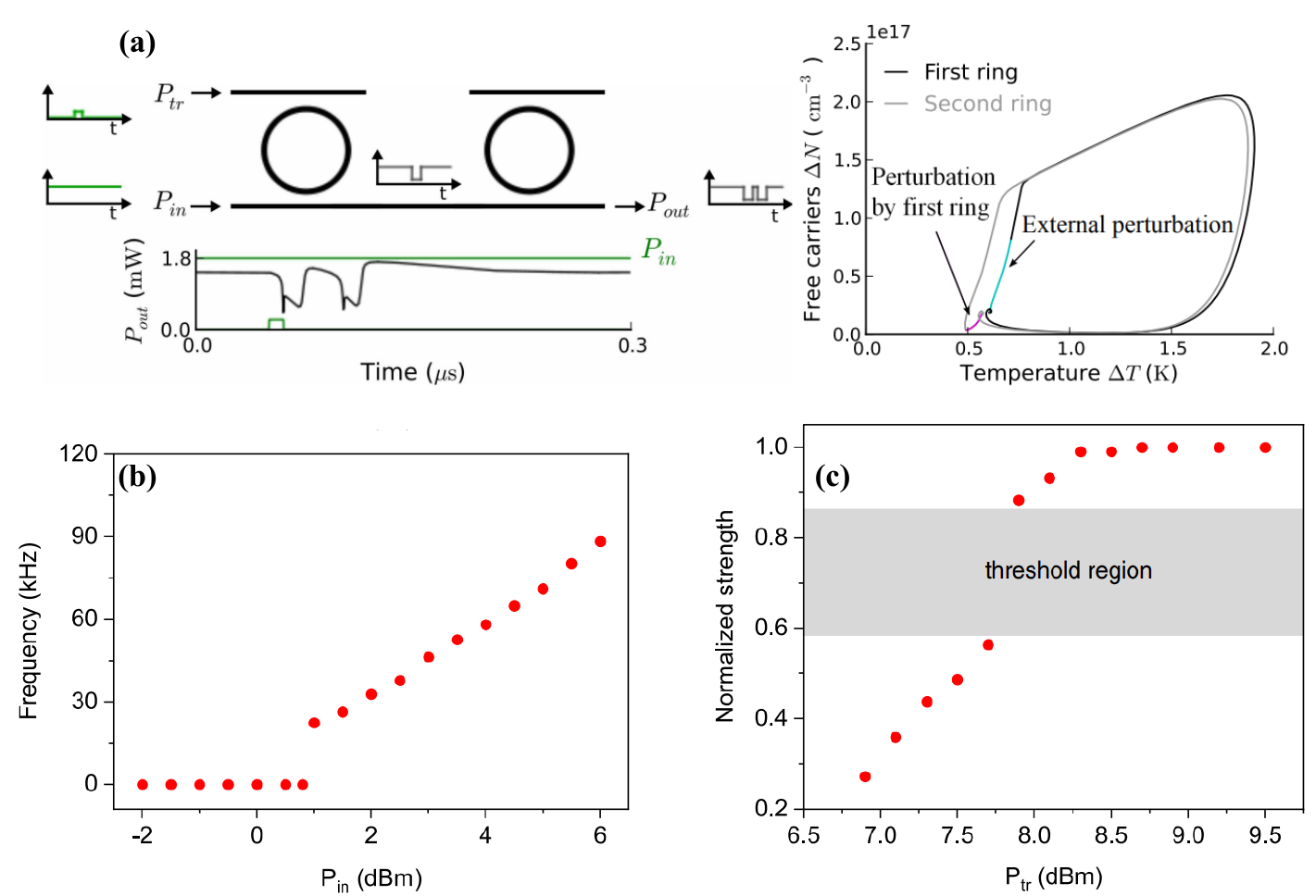}
	\caption{(a) on the left, a sketch of the system consisting of two MRRs in an add-and-drop configuration connected by a bus waveguide. Using a pump and probe approach, a self-pulsing oscillation can be induced by a small probe perturbation. The injection of the probe in the add port ensures the cascadability of the system. The corresponding phase portrait is shown on the right. Reprint by permission from \cite{van2012cascadable}. (b) Frequency of self-pulsing as a function of input power. (c) Normalized spike strength as a function of perturbation power. Reprint by permission from \cite{xiang2022all}.}
	\label{fig:ClassII}
\end{figure*}

In the self-pulsing region where the hysteresis cycle exists, i.e., the subcritical Andronov-Hopf bifurcation, the MRR is excitable when the input power is below, but close to, the limit cycle bifurcation. In this condition, a small perturbation takes the MRR from the resting state to a state characterized by a single self-pulsing oscillation, which then returns the system back to the initial resting state. There is no stable limit cycle that gives rise to continuous self-pulsing oscillations, and the MRR falls into class II neural excitability \cite{van2012cascadable}. This response is shown in Fig. \ref{fig:ClassII} (a) by means of a system consisting of two add and drop MRRs connected by the through port. The working condition is set by a constant pump laser with a power of 1.8 mW injected into the through port. The perturbation is generated by exciting the drop port via a probe laser consisting of a 10 ns rectangular pulse with a power of about 250 $\mu$W. Neglecting the backscattering phenomena (see subsection \ref{subsec:Back}), the perturbation affects only the first MRR, which exhibits a self-pulsing oscillation in the through port and then turns back to its rest state. This structure also demonstrates cascadability. That is, the possibility to excite another MRR by the output response of the previous one. The relative temporal phase evolution is shown in the phase portrait to the right of Fig. \ref{fig:ClassII} (a).   

Class II neural excitability was recently confirmed experimentally in a passive all-pass single-mode silicon MRR with a quality factor of 62000 \cite{xiang2022all}. To measure the relationship between self-pulsing frequency and input power, a CW pump beam excites the MRR at $\Delta \lambda = 20$ pm. As the input power varies, the period of self-pulsing is estimated from the resulting waveform output of the bus waveguide. The result is shown in Fig. \ref{fig:ClassII} (a). Self-pulsing is present only when the power is above a threshold of 0.8 dBm, otherwise the response remains constant in time. This discontinuity reveals class II excitability. In addition, it was measured the normalized negative spike strength as a function of the perturbation power. To this purpose, the pump beam is tuned at -20 pm from a resonance wavelength, while a probe beam is set at 10 pm from a resonance wavelength belonging to another FSR. The probe is a rectangular pulse with a duration of 15 ns that repeats at a frequency of 5 MHz. The negative pulses measured at the through port in response to the probe are normalized in strength of ``negative" spikes relative to the largest negative pulse measured. The result of normalized strength as a function of perturbation (probe) power is shown in Fig. \ref{fig:ClassII} (b). The presence of a sharp jump between about 7.7 dBm and 7.9 dBm combined with the linear change in subthreshold spikes is a further confirmation of class II resonate-and-fire neurons.

Other typical features of spiking neurons, namely refractory period, temporal integration, and inhibitory behavior, have also been demonstrated in the passive MRR \cite{xiang2022all}. The refractory period is the recovery time required for a neuron to be triggered after an excitation. It permits the neuron to relax to its resting state allowing the repeatability of the spiking response. Since this time determines the operating speed of a spiking ANN, its presence in the case of a passive MRR has been demonstrated and measured \cite{van2012cascadable,xiang2020all}. By using a double probe pulses in a pump and probe configuration where the pump has a constant negative detuning with respect to the resonance wavelength ($\Delta \lambda < 0$), it has been shown that the refractory time is of the order of magnitude of $\tau_{th}$. In fact, after an excitation, the MRR is sensitive to a new excitation only if its temperature decreases enough to reach the rest state. Therefore, the refractory time of a MRR can be estimated from the exponential drop of $\Delta T$ following an excitation pulse \cite{van2012cascadable}. However, it has been shown theoretically that the refractory time can be suppressed by increasing the power of probe pulses \cite{xiang2022all}. Using about 80 ns spacing between the probe pulses, a pump power of 1 mW with $\Delta \lambda = $20 pm and a probe peak power of about 0.6 mW, the MRR responds to the first excitation (probe pulse) but not to the second excitation. In contrast, by increasing the probe power to about 2 mW, the MRR responds to both pulses by exhibiting two single self-pulsing oscillations \cite{xiang2022all}. Consequently, the passive MRR can operate at speeds not limited by $\tau_{th}$ for high probe powers. 

Temporal integration of a neuron refers to the ability of the neuron to be excited and produce an output spiking due to the integration of a set of sufficiently close subthreshold pulses. On the other hand, inhibitory dynamics refers to the stopping of a spiking activity by a stimulus. Both features have been observed and experimentally demonstrated in the case of a passive MRR \cite{xiang2022all}. In addition, the spike-timing-dependent-plasticity, namely the biological process in which a precise timing of spikes adjusts the strength of the connections between neurons, has also been numerically demonstrated using a scheme based on MRRs \cite{mesaritakis2020micro}.   

\begin{figure*}[t!]
	\centering
	\includegraphics[width=1\textwidth]{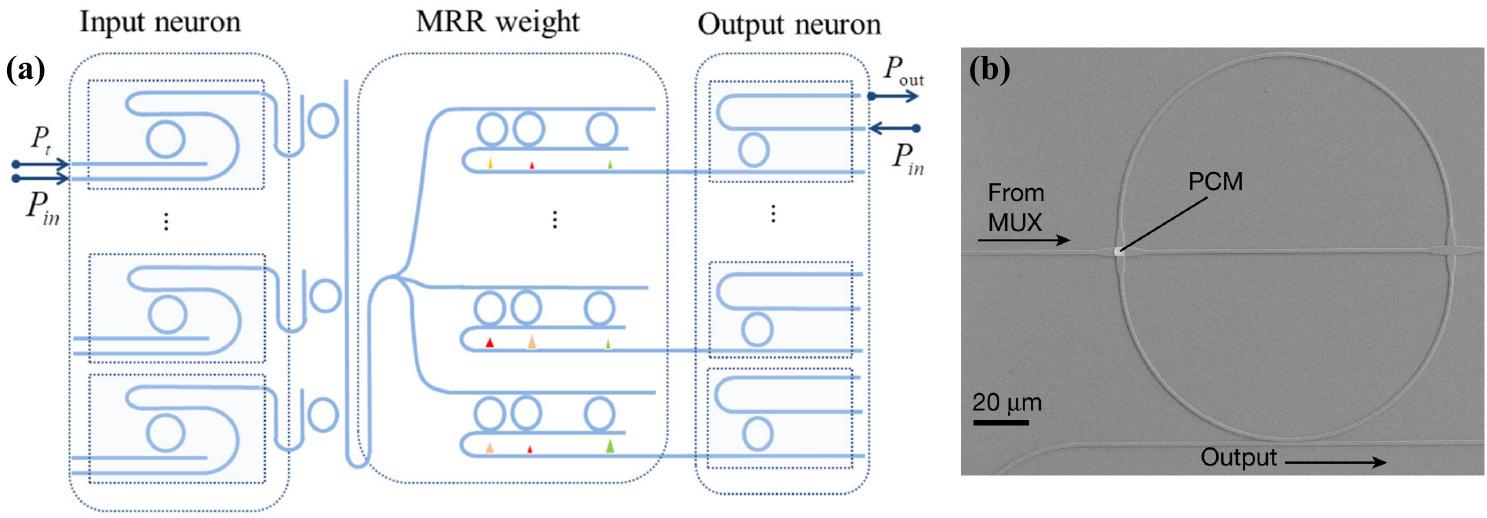}
	\caption{(a) Sketch of the all-optical neural spiking network based on the MRRs. Reprint by permission from \cite{han2022all}. (b) Scanning electron microscope image of the MRR used as the activation function. The PCM cell is placed on top of the waveguide crossing. Reprint by permission from \cite{feldmann2019all}.}
	\label{fig:Spiking}
\end{figure*}

Emulation of major neurocomputational properties paves the way for the creation of complex topological architectures of spiking ANNs with MRRs. They can be used at different scales of their temporal dynamics, from slow thermo-optic effects to faster processes such as free carrier dispersion or the instantaneous Kerr effect. However, only a few spiking ANN architectures based on the intrinsic nonlinearity of silicon have been proposed. An example is shown in Fig. \ref{fig:Spiking} (a) \cite{han2022all}. It consists of a layer of input neurons, a layer of weights, and a layer of output neurons. In the former, information is encoded in different spiking times at different wavelengths. It is then split into different channels via waveguides and sent to the weights layer. Here, the strength of the spikes generated by the input neurons can be adjusted by means of the coupling between the MRRs and the bus waveguides. The resulting signal is finally sent to the output neurons, which either fire or do not fire, producing the network response. The neurons are formed by MRRs in add and drop configuration and require a pump laser to set the initial condition to induce a spiking phenomenon, such as a self-pulsing oscillation. 

Noteworthy, there are other implementations of photonic spiking ANNs based on passive MRRs \cite{feldmann2019all} which implement a neurosynaptic system capable of supervised and unsupervised learning. In this spiking network  not based on silicon nonlinearities, MRRs are formed by a silicon ring closed by two waveguide crossings (Fig. \ref{fig:Spiking} (b)). A phase change material (PCM) cell is placed above the crossings. This allows changing the resonance condition and the propagation losses of the MRR by simply changing the PCM state between crystalline and amorphous. When the PCM cell is in the amorphous state, the optical input signal is not in resonance with the MRR. In contrast, when the PCM cell is in the crystalline state, the optical input signal couples to the ring and, thus, an output spike is observed. The implementation of a PCM cell as a tool to control the activation function has been previously studied and demonstrated in \cite{chakraborty2018toward}. These hybrid systems will be the subject of a detailed discussion in section \ref{sec:Hybrid}.

\subsection{Artificial neural network based on a spatial distribution of nodes}

In the following, we distinguish different implementations of MRRs within ANNs based on the quantity that is used to implement the network (i.e. how the neurons/nodes are distributed): the spatial, the temporal, and the wavelength domains. The spatial domain includes those implementations of ANNs whose dimensionality lies on multiple different physical nodes. The spatial arrangement of the nodes and their interconnections can lead to a variety of topologies. A first classification concerns feed-forward and recurrent ANNs, and is used here to present the different optical implementations. 

\begin{figure}[t]
\centering\includegraphics[scale=0.43]{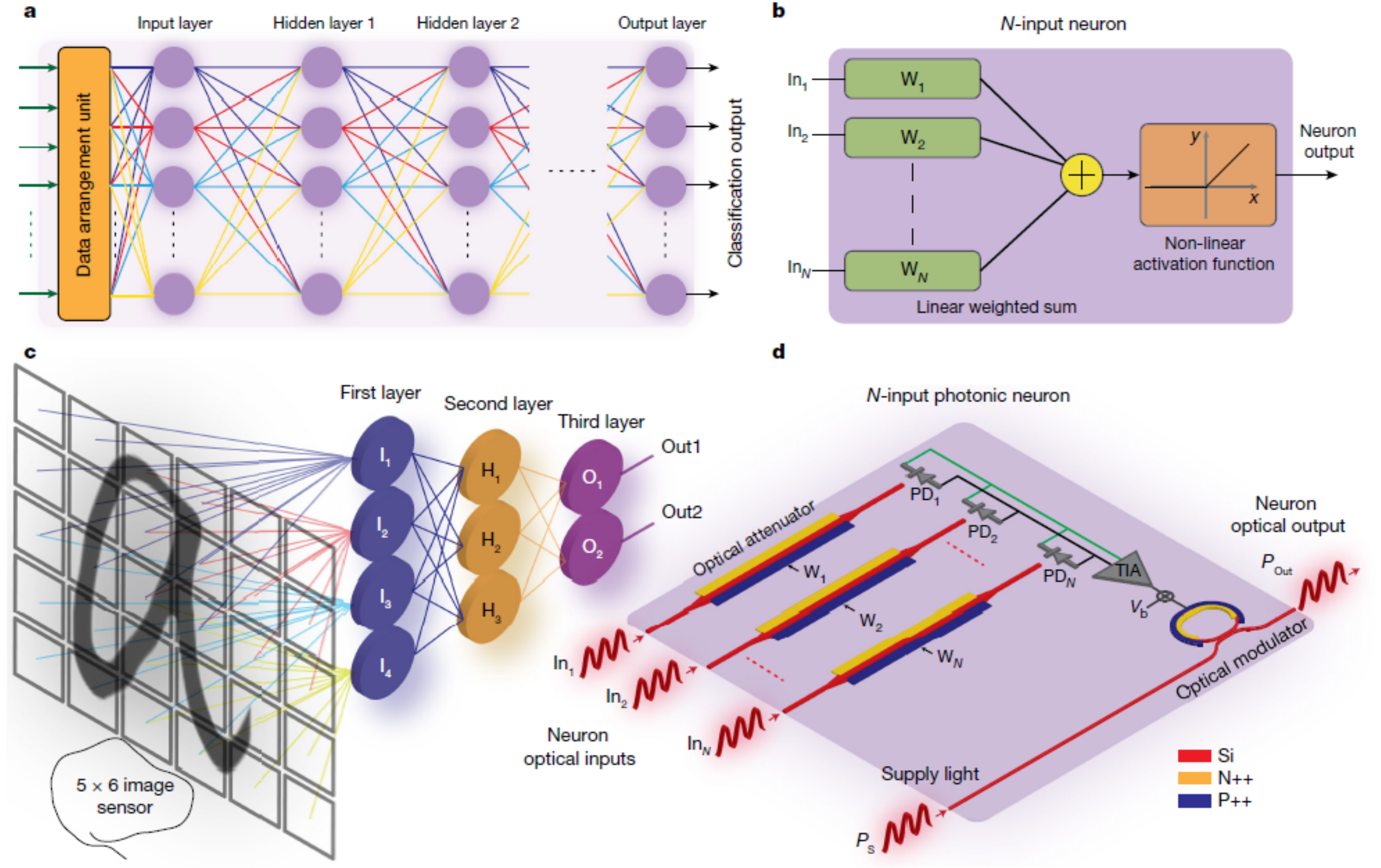}
\caption{a) Feed-forward architecture scheme. b) Artificial neuron scheme. c) The pixel values of an image to process are interpreted as state of the input layer neurons. d) Example of opto-electronic neuron realized in an integrated deep ANN. Reprint by permission from \cite{ashtiani2022chip}.}
\label{Fig_deepFFNN}
\end{figure}

In the feed-forward ANN scheme, see Fig. \ref{Fig_deepFFNN}(a), neurons are arranged in layers such that the data  can only flow in one direction, from the input to the output layers. When a large number of layers are present, the term "deep" ANN is sometimes used. Each neuron receives input from neurons of the previous layer, applies a nonlinear transformation to the weighted sum of these, and passes the result to neurons of the next layer (Fig. \ref{Fig_deepFFNN}(b)). The used nonlinear functions are different: binary step function, linear, sigmoid, tanh, ReLU, exponential linear unit, and other functions \cite{sharma2017activation}. The physical neuron response determines which one is used. The input layer enters the data (which represent the information to be processed) in the network and the last (output) layer provides the answer. A common benchmark for feed-forward ANNs is the recognition of handwritten letters, where the pixel values of an image representing a given letter are the information that, once serialized, feed the first layer of the network (Fig. \ref{Fig_deepFFNN}(c)). 

An example of a feed-forward photonic ANN, specifically designed for image classification, is reported in \cite{ashtiani2022chip}. Here, a laser beam opportunely collimated (beam diameter 870 $\mu m$) is shined on a letter drawn on a custom-made Plexiglas holding frame, and the resulting image is collected by a 5x6 matrix of grating couplers (150 $\mu m$ x 140 $\mu m$). Due to the high losses at this step (estimated at 41dB), the laser is amplified up to 63 mW. The grating signals are then routed to the first of a three layers network by means of waveguides, y-junction splitters and waveguide crossings. The layers are composed of four, three and two neurons respectively, and are fully connected. The neurons of the first layer are made by 500 $\mu m$ long waveguides with a p-i-n junction across. The input optical signals propagate in these waveguides whose transmission is weighted by the current injected into the forward bias p-i-n junctions. In this way, the trained weights are applied to the data. Then, the output optical signals are individually detected by silicon-germanium photodiodes (PD). The photodiode currents are summed, amplified and converted to a voltage by a transimpedance amplifier (TIA). The resulting voltage drives a forward bias pn-junction which is placed across the output MRR (Fig. \ref{Fig_deepFFNN}(d)). A CW optical signal is coupled to the MRR, and nonlinearly transmitted depending on the voltage applied to the pn-junction. The applied nonlinear transformation is a ReLU nonlinear function that is obtained by controlling the relative wavelength position of the MRR resonance with respect to the input CW signal. This represents an opto/electronic integration scheme which is possible due to the CMOS-compatibility of silicon photonics. Note also that all neurons in the network are externally powered by CW light, mitigating scalability issues induced by optical losses. This implementation is also the first demonstration of an end-to-end fully integrated feed-forward network, where external computation only supports the training phase. On the other hand, input pre-processing operations and the computation of the final response of the network are left to off-the-chip hardware. 

In \cite{ashtiani2022chip}, the off-the-chip sensing of the input data induces large total optical power losses of the feed-forward network. Another example of a feed-forward network where the input data are directly encoded on the PIC is discussed in \cite{biasi2022effect}. Here, all process steps from the encoding of the input information to data processing take place within the PIC. In particular, encoding and complex field weighting are performed by current-driven phase shifters. They consist of metal microheaters placed above the waveguides. The nonlinear activation function is provided by three MRRs that are thermally controlled to set the operating point.  Although the distance between the microheaters of the input layer and the MRRs is more than 800 $\rm \mu m$, a large thermal cross-talk was observed. The ANN uses the global heat generated by the encoding MRRs to reproduce the tasks. Consequently, the heat flow between the microheaters and the MRRs degrades the performance of the feed-forward ANN at the time scales of the thermal effects. This result has been explained by modeling the influence of ``local'' and ``global'' temperature variations in the nonlinear response of a MRR. The term ``local'' refers to the temperature of single optical components, such as the microheater and the MRR. The term ``global'' denotes the temperature of large regions of the entire PIC, including the substrate.

\begin{figure*}[t!]
	\centering
	\includegraphics[width=1\textwidth]{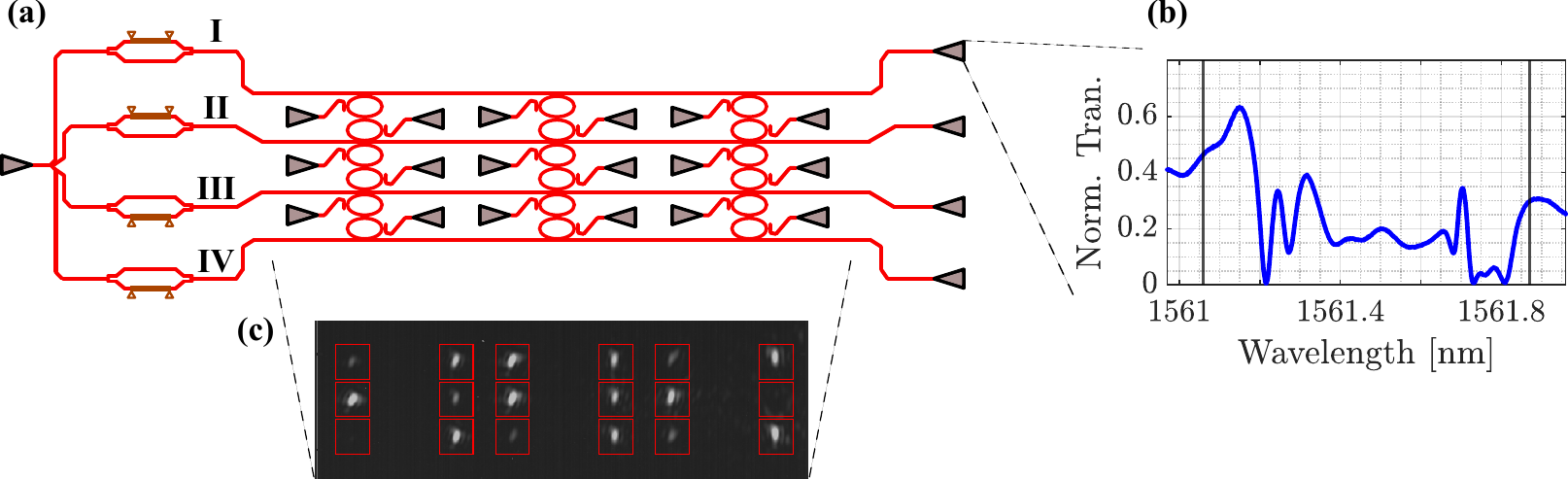}
	\caption{(a) Design of the PELM neural network. It consists of an input layer where the input data encoding takes place, a hidden layer consisting of an array of 18 microresonators coupled to grating couplers for scattered light monitoring, and a set of output grating couplers for response calibration. (b) Normalized transmittance as a function of the input wavelength for a response around 1561 nm, measured on the grating coupler indicated by the dashed black lines. (c) Example of the scattered light image acquired by an infrared camera. Reprint by permission from \cite{biasi2023array}.}
	\label{fig:PELM}
\end{figure*}

Although feed-forward neural networks are one of the most widely used deep-learning algorithms, they typically suffer from a time and energy consuming training procedure involving response optimization via a slow gradient descent algorithm \cite{rumelhart1986learning, van1998feed, lupo2021photonic}. A promising alternative is the Extreme Learning Machine (ELM) \cite{huang2006extreme, huang2011extreme}, namely, a feed-forward neural network consisting of a single hidden layer in which training occurs only in the output layer, while internal connections are random \cite{pierangeli2021photonic}. A recent experimental proof-of-concept demonstration of the implementation of a Photonic ELM (PELM) through integrated MRRs is shown in \cite{biasi2023array}. Here, the hidden layer of the PELM consists of an array of 18 MRRs, each one coupled to an output grating scatterer. Due to the linear propagation of the input optical fields in the MRR array and the nonlinear detection of the scattered light by a video camera, the input data are nonlinearly mapped into a scattered light image. Training is done offline by analyzing the recorded scattered light images with a linear classifier using a digital hardware. The network sketch is shown in Fig. \ref{fig:PELM} (a). The red lines represent the waveguides, while the triangles are the grating couplers. The information is encoded in the PIC at the input layer. This is formed by an input grating linked to four channels (I, II, III, IV) by a 1x4 multimode interferometer. The input CW optical signal is split in 4 and each input optical signal passes through a balanced Mach-Zehnder interferometer, which allows the amplitude of the signal to be modulated, encoding the input information. Then, the four optical data signals enter the MRRs matrix, which performs a weighted linear combination that maps the input space to a higher dimensional output space. The randomness of the weights is ensured by the stochastic fabrication errors on the MRR (MRR perimeters, bus waveguide/MRR gaps , MRR coupling coefficients). Figure \ref{fig:PELM} (b) shows the normalized transmittance as a function of the incident wavelength for a resonance around 1561 nm measured on the output grating, indicated by the dashed black lines. The presence of the fabrication errors induces a broadening of the MRR resonance to a band composed of different local minima. Figure \ref{fig:PELM} (c) shows the image of the light scattered by the MRRs captured by a camera for a given wavelength. The PELM has been tested by solving binary (logical operations) and analog tasks such as iris flower classification and banknote authentication.

\begin{figure}[t]
\centering\includegraphics[scale=0.45]{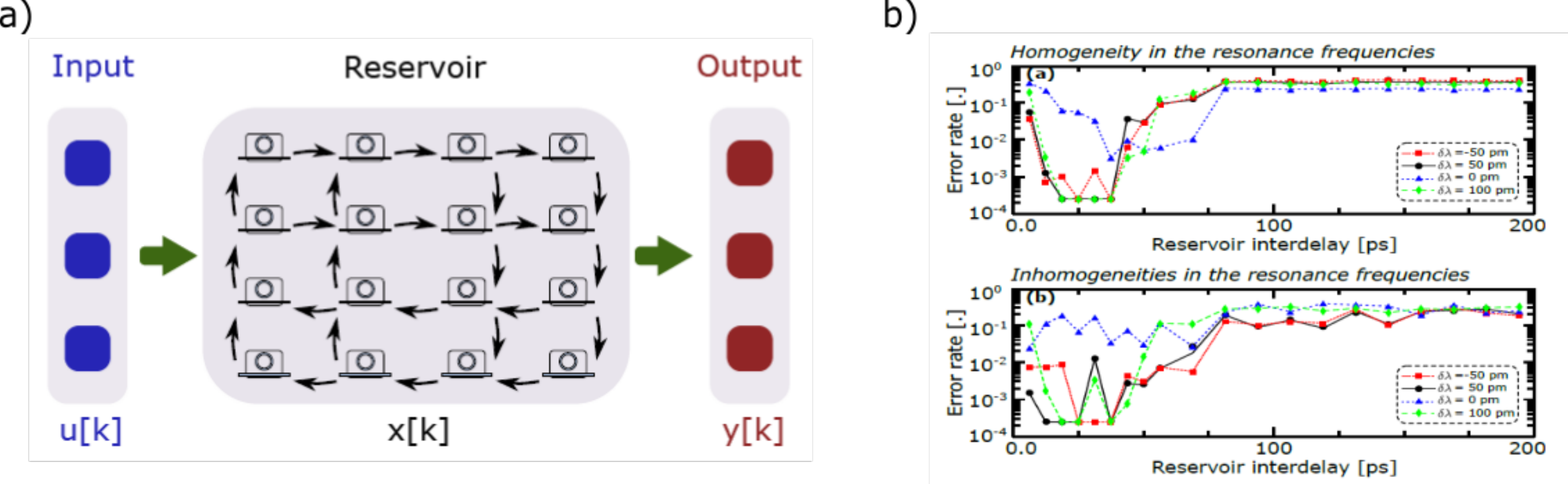}
\caption{a) Reservoir computing swirl architecture where silicon MRRs are the physical nonlinear nodes. The state of all the MRR is extracted to compute the output layer of the network. b) Simulation results obtained when testing the structure on the 1-bit delayed XOR task as a function of the reservoir interdelay between consecutive MRRs, in absence (up) and accounting (down) for resonance shifts induced by fabrication errors. Reprint by permission from \cite{denis2018all}.}
\label{Fig_RC}
\end{figure}

Unlike feed-forward ANNs, recurrent topologies allow for backward connections. As a result of recurrences, the network's state and its output are affected by past information still present inside the network. Thus, a small change in the topology may confer memory properties to the network, allowing it to solve memory-demanding tasks. On the other hand, the presence of recurrences complicates the training algorithms, resulting in a longer computation time. An approach to recurrent networks that simplifies their training is Reservoir Computing (RC) \cite{maass2002real}. Here, the recurrent network is essentially considered as a black box, and the only trainable connections are those projecting to a linear readout layer. An integrated implementation based on silicon MRRs is numerically investigated in \cite{denis2018all}, and schematized in Fig. \ref{Fig_RC}. The reservoir is realized with a 4x4 matrix of MRRs, connected in a swirl topology by waveguides with 3.0 dB/cm losses. The input layer, which introduces the data into the reservoir, is realized by simultaneously injecting the same input to all 16 nodes with random phase shifts. The output layer is computed numerically by linearly combining the detected states of all the MRRs. As a result, in this implementation, the connectivity matrix is given by splitting ratios, losses and random phase shifts, while the MRRs represent the physical nonlinear nodes of the network.      

The network is tested on the 1-bit delayed XOR task, a nonlinear boolean operation between every current injected bit and the previous one, whose output is estimated by the optical response of only the current one. Thus, the network is required to store 1 bit of past information, and to exhibit a proper nonlinearity for solving the task. The region of best performance is explored in a parameter space spanned by the input optical power (return-to-zero input signal), the detuning between the laser and MRR wavelengths, and the interconnect length between the MRRs. Figure \ref{Fig_RC} shows the obtained results in terms of the Bit Error Rate (BER), i.e. the rate of errors which measures how many bits are wrong with respect to the target value. Two cases are considered at a rate of 20 Gbps: a homogeneous case where all MRRs have the same resonance, and a more realistic inhomogeneous scenario where fabrication errors induce resonance deviations that follow a Gaussian distribution with a standard deviation of 10 pm. While these results demonstrate the robustness of the network against fabrication errors, they also highlight the need to detune the input optical signal with respect to the MRR resonance wavelength. Indeed, when the light is closer to the resonance condition, more optical power is absorbed by the MRRs and less power spreads along the network. Also, the interconnect lengths must be optimized to spread the information, on the base of the input injection rate. From this last observation, it can be argued that recursive connections within a network do not necessarily lead to the desired memory effects. In this implementation on the relation between the input modulation rate and the interconnection delay at the data wavelength. 

Similar RC architectures based on matrices of InGaAsP/InP MRRs are studied numerically in \cite{mesaritakis2013micro} and \cite{mesaritakis2015high}. The MRRs are now interconnected using a rainfall topology, where each MRR input is supplied by the drop signal of the previous one. Additional feedback loops connect the last column of MRRs to previous columns, further extending the fading memory of the system. All the connection delays and strengths are chosen randomly, provided that the optical power is attenuated when propagating between two MRRs. This is to suppress potential chaotic behavior that opposes the deterministic nature of the computation. This time, the optical information is fed into the input port of only one ring in the first column, while the output signals are provided by the drop ports of all the MRRs in the last column. The detected signals are then fed to an electronic perceptron whose weights are trained to optimize the task performance. Matrices of 25 (5x5) and 36 (6x6) InGaAsP/InP MRRs are investigated, thus having 5 and 6 optical output signals with a maximum delay of 88 ps and 1.2 ns, respectively. An additional EDFA is included to amplify the optical signal at the input of the reservoir and also to tune the OSNR (optical signal-to-noise ratio) and study its effect on the network performance. Relying on the MRR nonlinearities triggered by TPA and Kerr effects (with $\tau_{fc}=100$ ns), the 5x5 reservoir is applied in 3-bit and 8-bit pattern recognition tasks with classification errors of 0.1\% (at 40 Gbps) and 0.5\% (at 160 Gbps), while the 6x6 reservoir is applied to identify images acquired by the dispersive Fourier technique. Note that in all cases, a time interval is allocated between consecutive inputs so that the initial state of the reservoir is always restored.  

For ANNs using MRRs as activation nodes, the distinction between feed-forward and recursive networks may not be well defined. In fact, the presence of backscattering due to surface roughness (see section \ref {subsec:Back}) can induce recursive phenomena even in the case of simple feed-forward networks. From a manufacturing point of view, the quality factor should be defined as a balance between the power to activate nonlinear phenomena and the way in which the field inside the MRR is affected by surface roughness. This may be one of the reasons for the few experimental implementations of pure feed-forward networks based on the nonlinear response of passive MRRs.

\subsection{Artificial neural network based on a time distribution of nodes}

An alternative way to realize a recurrent network for reservoir computing (RC) is time multiplexing the input in the dynamics of a single nonlinear node. This possibility, originally introduced in \cite{appeltant2011information}, is attractive from an experimental point of view because it simplifies the physical network to a single node. On the other hand, a virtual network is generated in the time domain, since the input at different times is coupled by the nonlinear dynamics of the node. A MRR represents one of the possible single node candidates. 

\begin{figure}[t!]
\centering\includegraphics[scale=0.41]{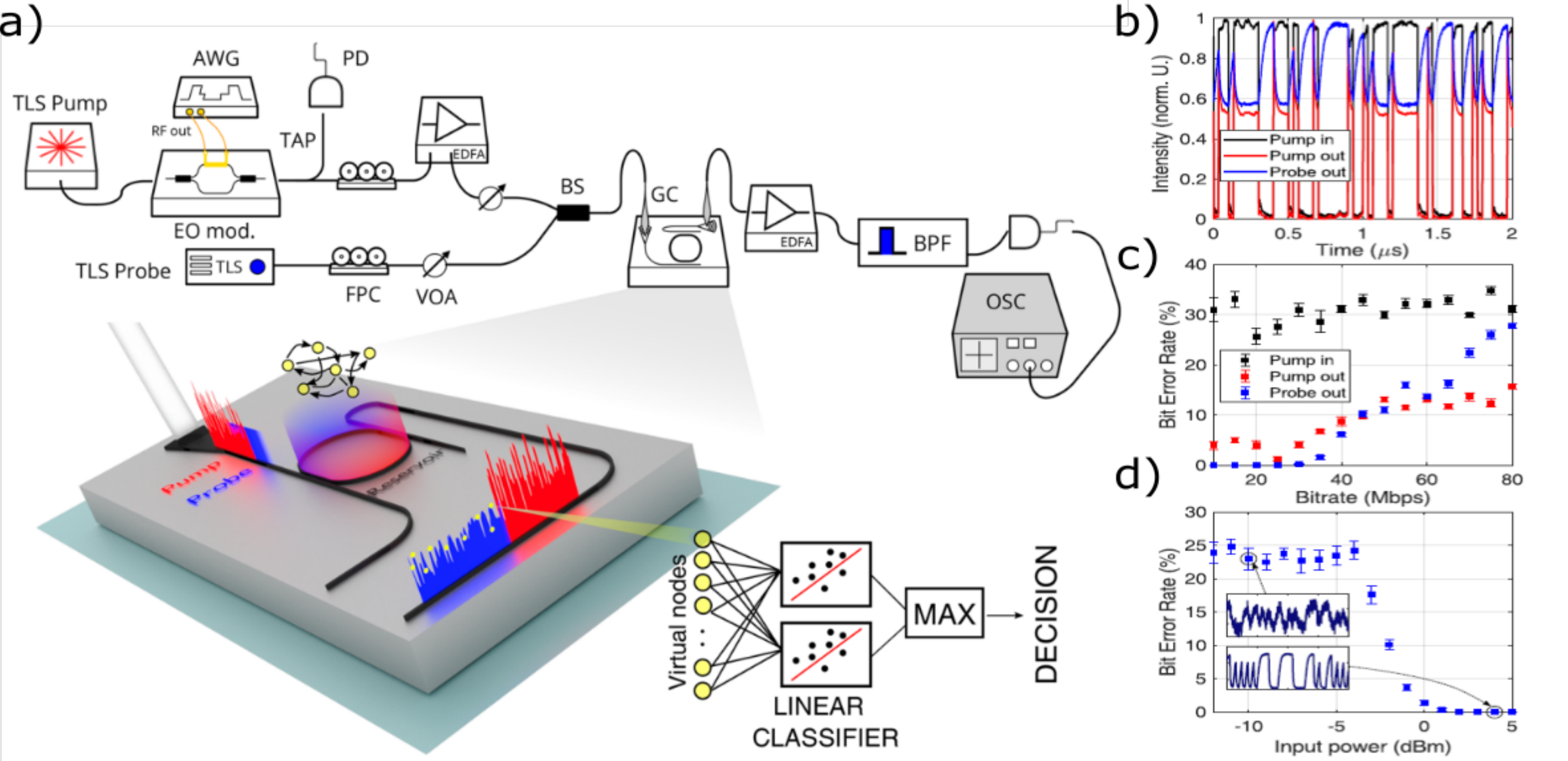}
\caption{Silicon MRR-based reservoir computing tested on both the 1-bit delayed XOR and the iris flower recognition tasks. The reservoir states are time-multiplexed in the MRR drop output signal, from which the virtual nodes are sampled and used to compute the output of the network. b) A pump and probe approach avoids null-signal responses, which otherwise would degrade the performance on tasks. c) - d)  Results on the 1-bit delayed XOR task as a function of the input bitrate and power, while using 3 virtual nodes. Reprint by permission from \cite{borghi2021reservoir}.}
\label{Fig_Borghi}
\end{figure}
A first experimental investigation using a silicon MRR as a single node for RC is reported in \cite{borghi2021reservoir}. Here, the MRR is designed in the add-drop configuration. It receives the input data at the input port, and provides its dynamic response at the drop one (see Fig. \ref{Fig_Borghi}). For each input data, the MRR drop port transmission is detected and sampled offline at specific times (virtual nodes) to provide the state of the folded reservoir. Then, the output layer is computed by a weighted linear combination of the virtual node values. The reservoir training consists of determining the weights to solve a given task. They are estimated by offline regularized least squares (ridge regression). By using the free carriers and thermal nonlinearities of the MRR, discussed in section \ref{subsec:NonLinear}, both the 1-bit delayed XOR task and the iris species recognition task are solved. Note that, the input data bit is nonlinearly mapped via the dynamics of the physical node into a higher dimensional space (given by the number of virtual nodes). This requires that the time duration of the input bit is enough long to sense the nonlinearity of the node. For example, when computing the 1-bit delayed XOR task, each bit is encoded at a bit rate close to the free carriers lifetime (45 ns, $\sim$20 MHz). This is highlighted in Fig. \ref{Fig_Borghi}(c), where the performance on the task is reported as a function of the input bitrate. 
This implementation also teaches a clever method to avoid having null optical output signals as a consequence of null optical input signals. Indeed, a pump and probe approach is used, where the data is encoded into the amplitude of a pump signal and the MRR response is imprinted onto a second CW probe signal at a different resonance wavelength. Unlike the output pump signal, which is noisy for 0-bit, the probe always carries a non-zero signal that is affected by the MRR nonlinearities triggered by the pump signal (see Fig. \ref{Fig_Borghi}(b)). The results show that when the virtual nodes are sampled on the input pump the task is not solved, while improvements emerge by using the output pump signal, due to the MRR nonlinear transformation. Nevertheless, the best performance is achieved using the output probe signal, where the best result is found at 25 Mbps for an input pump power larger than 2 dBm. The ability to solve the 1-bit delayed XOR demonstrates the presence of memory and nonlinearity in the reservoir, caused by the inertia of the free carriers nonlinearity itself. For example, the free carriers generated by an input equal to 1 are still relaxing when the next input arrives, thus affecting future-bit responses. 

Further experimental studies on linear and nonlinear memory demanding tasks with a single MRR as a reservoir are performed in \cite{Bazzanella2022}. The same approach as in \cite{borghi2021reservoir} has been used here, but with a single pump laser and a MRR having free carrier dynamics with a relaxation time of few ns. The logical XOR and AND operations were performed between the current bit and more than one past bits. From the results emerges that the free carrier-based nonlinear memory allows to store up to two bits in the MRR state. This memory allows solving the linear delayed AND task considering up to two bits in the past, but only the 1-bit delayed XOR task. The XOR task is indeed more complicated than the linear AND, since both memory and nonlinearity need to be used, and 2-bits memory turns out to be insufficient. To achieve an extended memory, one can vary the single node topology by considering more than one MRRs, or by adding delay lines. For example, in \cite{donati2022MRR} a silicon MRR coupled to an external optical feedback is numerically studied as a single node for RC. The system is here applied on time-series benchmark tasks, and shows that the external feedback is beneficial not only for extending the memory of the system, but also as an additional tool for tuning the MRR nonlinearity. Note also that while the feedback memory is provided by an optical signal, the free carriers (and thermal) memory is stored in the MRR state.

It should be emphasized that in the case of off-the-chip data encoding, the temporal approach to virtual nodes requires the analysis of the optical input signal and of the ANN output signal \cite{Bazzanella2022}. In fact, the use of an arbitrary waveform generator coupled to an optical modulator to generate the optical bit sequence induces a nonlinear transformation of the digital data that alone enables the readout regressor to solve the considered tasks.

\subsection{Artificial neural network based on a wavelength distribution of nodes}

One key advantage of photonics is the inherent parallelism it enables. Signals of several distinct wavelengths can propagate along the same physical medium, carrying simultaneous data. This opens the possibility to encode the neurons on different wavelengths, which contributes to increase the dimensionality and the bandwidth of photonic neuromorphic ANNs. The versatility of MRRs also finds applications in this context, where they can even serve as sources of different wavelength channels. 
\begin{figure}[t]
\centering\includegraphics[scale=1]{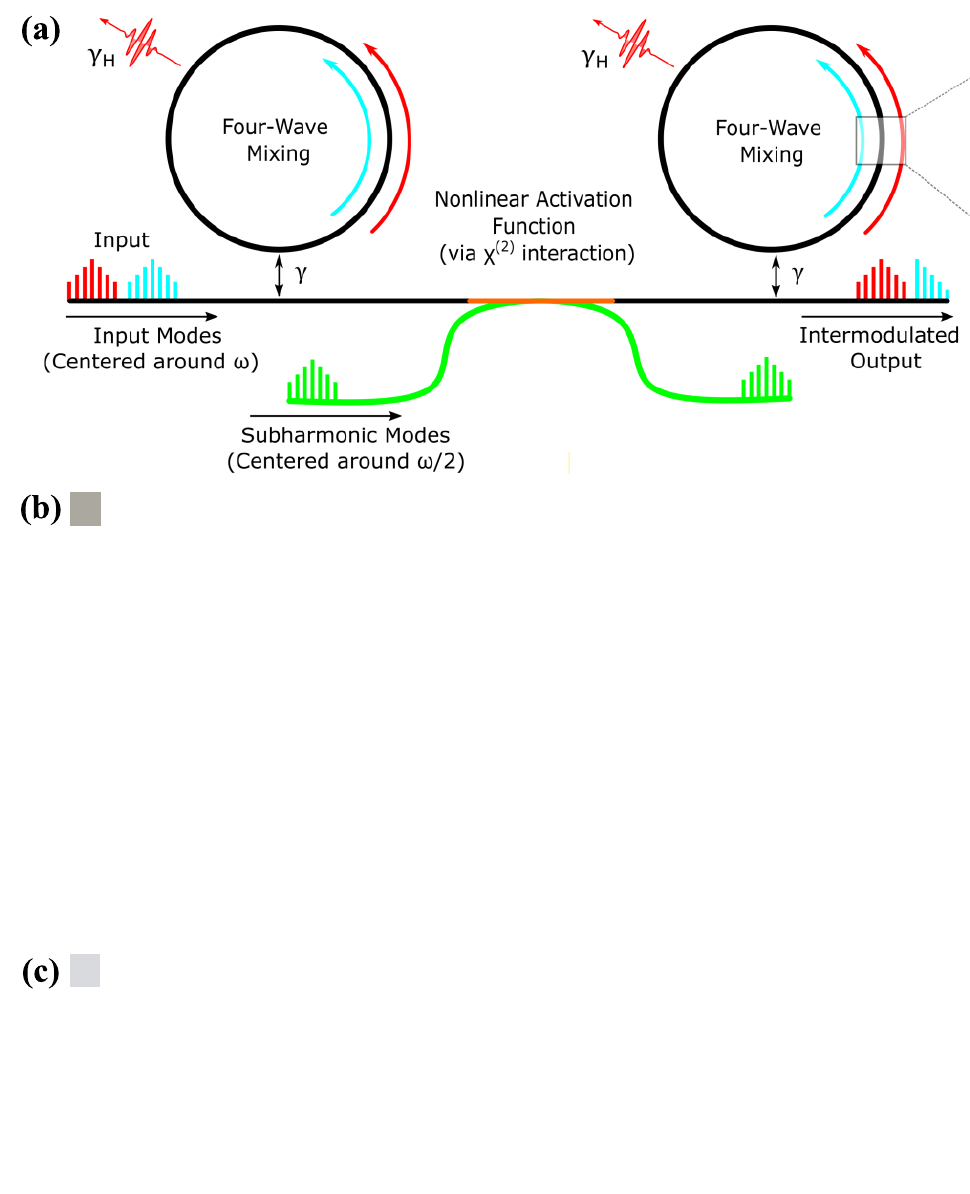}
\caption{(a) Sketch of the two MRRs coupled to a bus waveguide. The matrix-matrix multiplication occurs in the MRRs via four wave mixing (FWM), while the nonlinear activation function is provided by a $\chi^{2}$ medium on the bus waveguide. Reprint by permission from \cite{basani2022all}. (b) Silica MRR as a source of soliton crystal microcombs. (c) Experimental implementation of a photonic perceptron processing in parallel multiple wavelengths generated by an integrated silica MRR. Reprint by permission from \cite{xu2020photonic}.}
\label{Fig_MicroComb}
\end{figure}

The ability to take advantage of different wavelengths combined with the nonlinearity of the $\chi^{3}$ medium, led to the use of a passive MRR to induce matrix-vector and matrix-matrix multiplications through a Four Wave Mixing (FWM) process. In a stimulated degenerate FWM process, two degenerate photons from a pump laser are mixed with a photon from a probe laser to produce an idler photon \cite{borghi2017nonlinear, boyd2020nonlinear}. Such a process is efficient if energy and momentum are conserved. Conservation of momentum implies phase matching of the three interacting and propagating waves \cite{manna2016stimulated}. An all-optical ANN architecture that exploits only nonlinear processes is proposed in \cite{basani2022all, Basani:22}. Specifically, information is encoded in the complex amplitude of wavelength modes that are intermodulated between multiple wavelength modes by a strong pump power due to FWM. This process takes place inside a multimode MRR (see Fig. \ref{Fig_MicroComb} (a)) where a pump and probe lasers are simultaneously injected. Here, the probe beams are the input of the neurons. On the other hand, the nonlinear activation function is given by the interaction of the light exiting the MRR with a medium exhibiting second-order nonlinearity $\chi^{2}$, such as lithium niobate. This is achieved by interacting the outgoing modes of the MRR with subharmonic pumping modes. Such an activation part, as shown in Fig. \ref{Fig_MicroComb} (a), is placed in the output bus waveguide of the MRR. The ability to cascade the MRR and the activation function through the bus waveguide also allows the dimensionality of the network to be increased spatially. As a result, this implementation can be considered as a wavelength-space approach. It benefits from the encoding in different wavelength modes as well as from the integrability of the MRR and the activation function. A remarkable property of this network is that the operating speed is directly proportional to the power of the pumps which yield billions of matrix multiplications per second at tens of mW dissipation rates.

Based on Kerr nonlinear effects occurring within the MRR, it is possible to activate parametric wavelength conversion processes that, if cascaded, can lead to coherent and deterministic multi wavelength generations such as in soliton crystals \cite{cole2017soliton}. Importantly, the FSR of the MRR determines the minimum achievable spectral spacing as the FSR, and consequently the density of wavelength channels. An integrated MRR that achieves a record low FSR is reported in Fig. \ref{Fig_MicroComb}(b). It is fabricated in a CMOS-compatible silica platform, with a radius of $\approx 592$ $\mu m$, a Q-factor of $\approx 1.5$ million and a FSR of 48.9 GHz ($\approx 0.4$ nm). When suitably pumped, the  microring generates a deterministic soliton microcomb. The MRR can be used to feed the wavelength channels of a photonics ANN. In \cite{cole2017soliton}, the 49 different wavelengths generated by this MRR are used to provide the input for an optical (bulky) 49-input neurons perceptron, whose scheme is shown in Fig. \ref{Fig_MicroComb}(c). Two optical spectral shapers are consecutively applied to flat the microcomb power spectra and then set the desired trained weights, respectively. Wavelength channels with higher power reflect a stronger synapse, and vice versa. To emulate a perceptron, the value of the input nodes are multiplexed in time, and then modulated by a single electro-optical converter. As a result, the same input node is loaded at the same time on all the wavelength channels. To align in time the correct information to the perceptron, each wavelength channel is precisely delayed via the chromatic dispersion of a single-mode fiber. The multiplexed and time-aligned signals are then detected to provide the sequence of weighted input sums of the perceptron. This implementation illustrates a combination of time and wavelength multiplexing techniques.
An upgrade of the system relies on 90 wavelength channels over 36 nm at 1550 nm to realize a photonic convolutional accelerator \cite{xu202111}. Here, different convolutional kernels are processed simultaneously via dedicated subgroups of microcomb lines. The input injection scheme follows that of the perceptron, thus using a single E/O modulator and delaying appropriately each wavelength channel, while at the output an additional stage is operated to demultiplex the different kernels before separately detecting them. In this way, a number of physical output channels equal to the number of kernels are implemented, to enrich the computational power of the system. This second example illustrates the potential of time-wavelength-spatial multiplexing techniques combined together to achieve remarkably high 11 TOPS (trillions of operations per second). 

\section{Silicon microresonators as weight bank} \label{sec:weight_bank}
\begin{figure}[t]
\centering\includegraphics[scale=0.30]{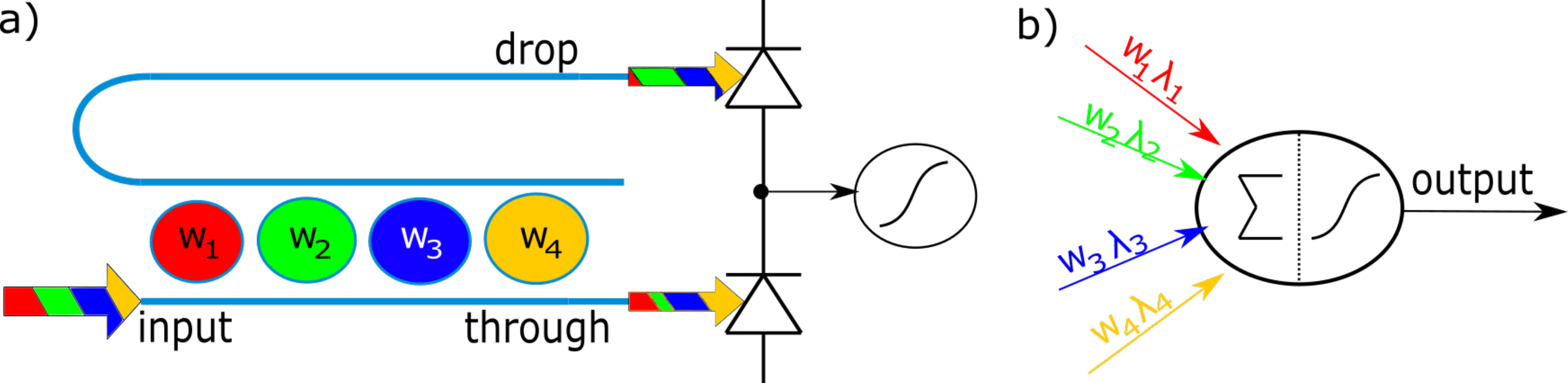}
\caption{a) A weight bank composed of an array of MRRs in an add-drop filter configuration. The structure is specifically designed to take advantage of wavelength-division-multiplexing (WDM), and applies specific weights ($w_i$) to the incoming wavelength channels ($\lambda_i$) by tuning the transmission of the MRRs. A balanced photodetector integrates the through and drop weighted signals, and enables for both positive and negative weights. b) Artificial/biological neuron representation, where input values are firstly weighted, then integrated and finally used as argument for the neuron nonlinear activation function. Differently from a biological neuron where a multidimensional input requires several physical channels, photonics equipped by WDM can employ only one. }
\label{Fig_weightBank}
\end{figure}

In the previous sections, several ways to exploit the nonlinearity of a silicon MRR both as neuron activation function and multi wavelengths source in photonic neuromorphic hardware have been described. On the other hand, alternative approaches use MRR linear filtering capabilities to encode the weights to an optical signal. A MRR placed at the input port of a photonic neuron is indeed able to transmit only a portion of the incoming signal, according to Eq. \ref{eq:Et-Ed}. Its transmission depends in particular on the detuning between the input and the resonance wavelengths. Whenever the detuning can be externally controlled by shifting the MRR resonance wavelength, thus changing its transmission, an analog and trainable weight can be obtained by means of a MRR. This concept is particularly useful when the data are encoded in several multiplexed wavelength channels. In this case, it is natural to deliver the weight operation to an extended set of MRRs, the weight bank, still placed at the input of a photonic neuron. In doing so, each wavelength channel can be uniquely weighted, and the intrinsic parallelism of photonic hardware can be exploited. In a common weight bank geometry (Fig. \ref{Fig_weightBank}), a set of MRRs shares the same input and output bus waveguides. Multiplexed wavelength signals enter the input port, are filtered by the weight bank, and their resulting two weighted versions are transmitted to the through and drop ports, where a balanced photodiode finalizes the optical weighted addition phase. The generated photocurrent encodes in this way the input weighted sum of the photonic neuron. It is important to note that the use of a  balanced photodiode allows to both apply positive and negative weights in the MRR weight bank. Indeed, the weight will be zero whenever the MRR equally splits the energy between the two ports, and positive or negative whenever an unbalance is present \cite{tait2015balanced}. 

The range of weight values that can be applied depends on the extinction ratio of each MRR. Ideally, the goal is to obtain the maximum extinction ratio, by having total energy transfer to the drop port when the incoming wavelength is perfectly in resonance, and total energy transfer to the through port when it is out of resonance. Designing the MRR in critical coupling allows to best approach this situation \cite{biasi2019time}, with intrinsic limitations induced by the MRR internal losses. Designing a weight bank with MRRs having a slightly different radius allows separating the individual resonance spectra. When the resonance spectra are completely separated, i.e. when there is no optical cross-talk, the device is described as an ensemble of independent MRRs, each one applying a weight to a specific wavelength channel. The radius also affects the FSR, i.e. the wavelength interval between consecutive resonances of one MRR (see section \ref{sec:intro}), which is the final critical limiting factor for the number of wavelength channels that the weight-bank can access and weight.\\
Because of fabrication errors, every MRR will be realized slightly different with respect to nominal parameters. Additionally, environmental changes, like temperature variations, and thermal and electrical cross-talks between the MRRs, also disturb the weight bank. As a result, precise calibration and control strategies have been developed. The latest progresses rely on appropriate sensing elements placed at each MRR site, to realize integrated feedback controls, aiming to set and maintain a desired weight \cite{de2022design}. For example, such a sensor can be realized by embedding a heater in the MRR waveguide, by lightly n-type doping \cite{tait2018feedback}. In this case, when an electric current is applied, a variation in the applied voltage can be sensed whenever light circulates in the MRR. In fact, donor-induced extra losses in the waveguide will produce new electron-hole pairs, which in turn lower the conductance. Note that the same sensor can also be used to tune the weight of the MRR, by heating and thus shifting its resonance wavelength, with precision up to 7 bits \cite{huang2020demonstration}, and up to 9 bits when the control extends to noise sources other than the MRRs \cite{zhang2022silicon}. Thermal stabilization of the PIC is an additional solution to balance temperature variations in the environment. 

A first proposal of an integrated photonic network implementing MRR weight banks was suggested in \cite{tait2014broadcast}, in what is known as the ``broadcast and weight" protocol. 
\begin{figure}[t]
\centering\includegraphics[scale=0.40]{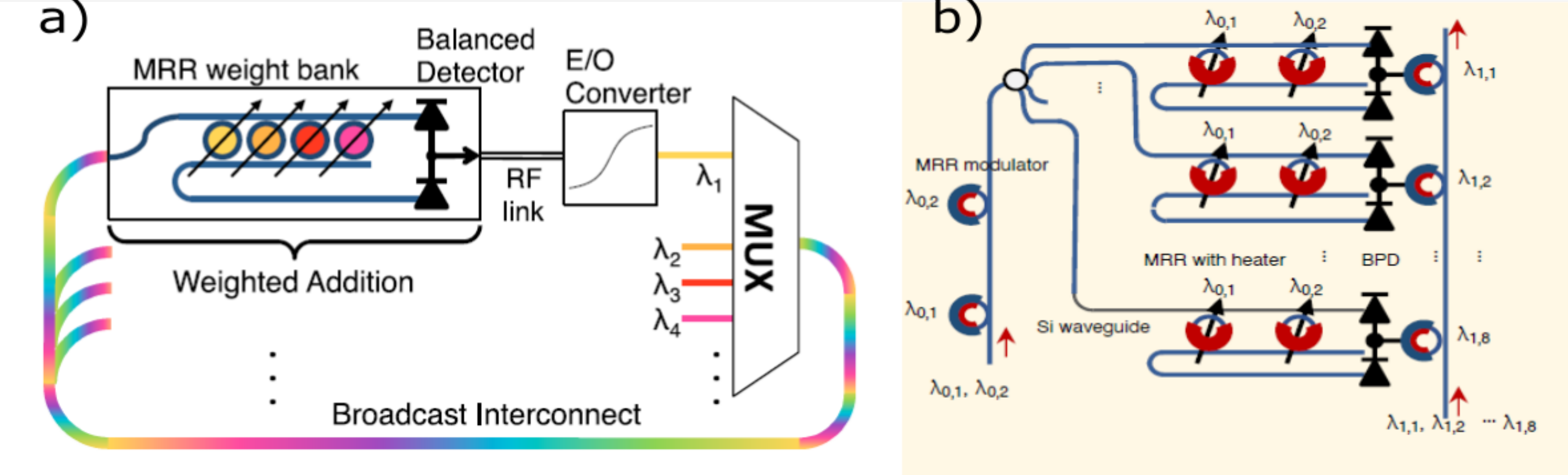}
\caption{a) Photonics implementation of ANN adopting the broadcast-and-weight architecture. A broadcast interconnect drives the signal to the photonics neurons. A MRR-based weight bank applies the input weights to the incoming signal, producing photocurrent that an electro-optical (E/O) converter nonlinearly imprints on a specific wavelength carrier, which is finally multiplexed into the broadcast interconnect. Reprint by permission from \cite{tait2016MRR} b) Representation of hidden layers in a photonics implementation of a feed-forward ANN model, based on broadcast-and-weight architecture. The output MRRs of the precedent layer j, apply a nonlinear electro-optic conversion on two CW wavelength channels multiplexed on a single waveguide ($\lambda_{j,1}, \lambda_{j,2}$). The optical information then spreads to the photonic neurons of the next layer, whose input weights are actuated via MRR-based weight banks. Reprint by permission from \cite{huang2021silicon}.}
\label{Fig_FFNN}
\end{figure}
The idea, schematically represented in Fig. \ref{Fig_FFNN}(a), is to multiplex N different wavelengths in a single common channel that brings the signals to the photonic neurons and at the same time collects their optical responses. The information is thus continuously reused by the network. Each photonic neuron is equipped with an N-MRR weight bank, which independently acts on each input wavelength channel, followed by an electro-optical converter. At this stage, the photocurrent generated by the balanced photodiode is nonlinearly imprinted on an output light signal at one specific wavelength channel, which is in turn finally multiplexed in the common channel. Choosing a different light-generation device, like an optically injected laser, is also possible but at the cost of rethinking the input sum stage, since it removes the advantage of using a balanced photodetector for generating the input weighted current \cite{de2019machine}. Considering electro-optical conversions, they can be implemented in several ways. For example, the balanced photodiode current can be delivered to a laser close to the lasing threshold, and eventually trigger an optical spike, or it can drive a Mach Zehnder modulator fed by a CW laser signal \cite{tait2017neuromorphic}. In a recent implementation, the electro-optical conversion is realized through a further MRR equipped with a pn-junction \cite{huang2021silicon}. In this case, the photocurrent signal is combined with a forward bias current to modulate the MRR transmission via free carrier injection. This approach is used in a broadcast-and-weight protocol to realize a feed-forward ANN model, having in particular two hidden layers as schematized in Fig. \ref{Fig_FFNN}(b). The hidden layers are composed respectively by two and eight neurons, and each neuron is provided by two MRRs that weight the two input wavelengths carrying the data. Interestingly, at the end of each layer, the two multiplexed wavelength carriers get restored by injection of a CW optical signal. The output signal from the MRR modulators of the previous layer is imprinted by the nonlinear electro-optic conversion on this signal, which feeds then the neurons of the next layer. The ANN has proven to be successful for telecom applications where the nonlinear effects in optical fiber transmissions can be mitigated. \\
\begin{figure}[t]
\centering\includegraphics[scale=0.5]{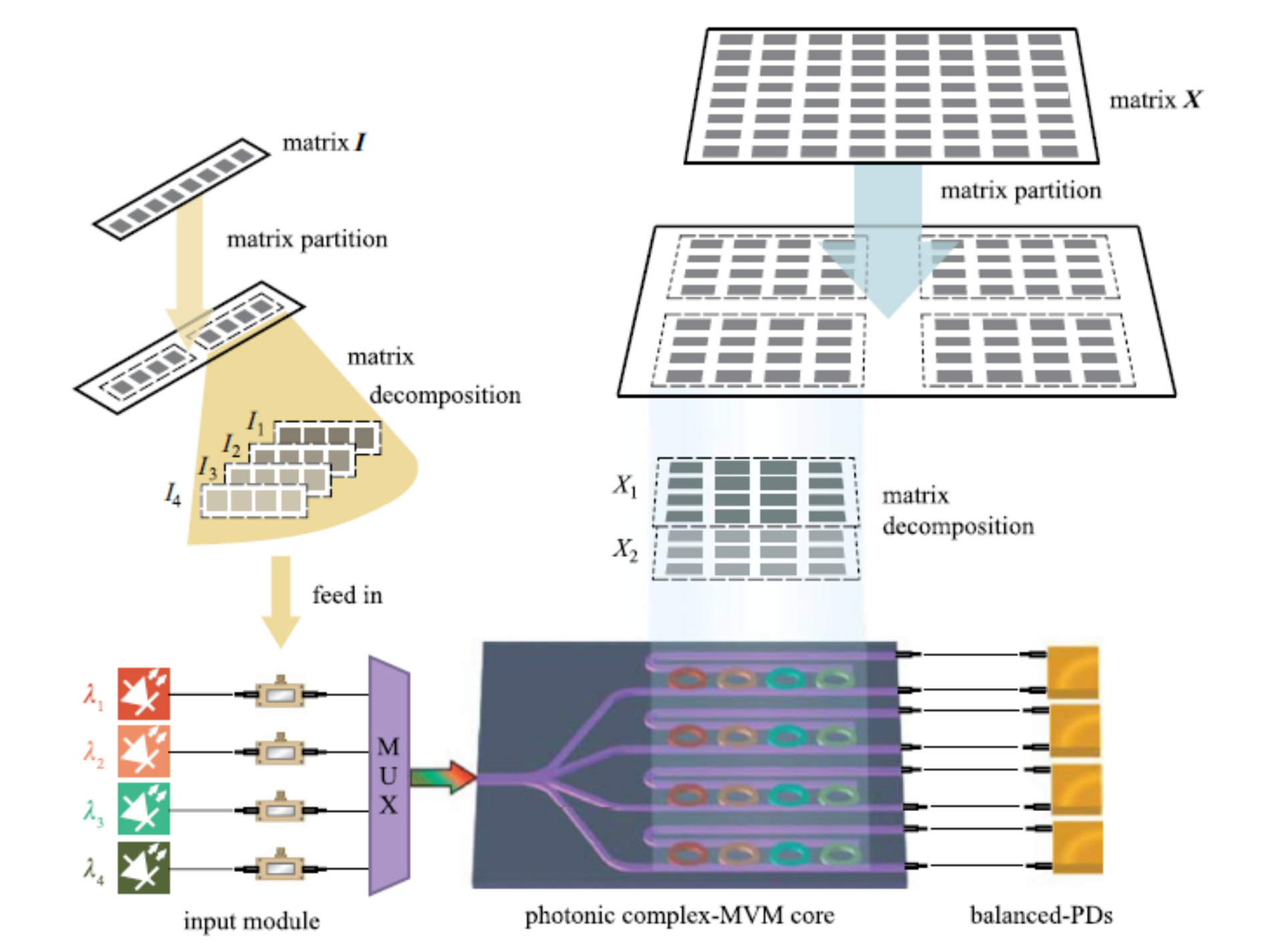}
\caption{A $4\times 4$ MRR-based weight bank applied as a tensor process unit.  Specifically, a matrix-vector multiplication $O=XI$ is operated, where $X$ is encoded by the $4\times 4$ weight bank and the $4\times 1$ input $I$ is amplitude encoded onto 4 corresponding wavelength carriers (($\lambda_1$, $\lambda_2$, $\lambda_3$, $\lambda_4$)). The operation can extend to the full complex domain by matrix partition and decomposition. Reprint by permission from \cite{cheng2022small}. }
\label{Fig_TPU}
\end{figure}
Microring-based weight banks are promising also as TPUs, i.e. an ASIC (application-specific integrated circuit) developed to efficiently perform matrix multiplications. An example of photonic TPU employing MRR-based weight banks is reported in \cite{cheng2022small}. Here a Matrix-Vector Multiplication (MVU) operation $O=XI$ is performed, being $O$ and $I$ respectively the output and input m-dimensional vectors, and $X$ an $m\times m$ matrix. In the experimental implementation, $X$ has dimensionality $4\times 4$ and is realized via 4 weight banks, each one composed of an array of 4 MRRs. Thanks to the combination with a balanced photodetector, as previously discussed, the matrix values (weights) can also be set negative, resulting in a -1 to 1 continuous weight range. As the MRR transfer function does not depend on the phase of the optical input signal, the input vector $I$ is loaded onto 4 optical signals having different wavelengths ($\lambda_1$, $\lambda_2$, $\lambda_3$, $\lambda_4$) by independent intensity modulators. By using matrix partition and decomposition (see Fig. \ref{Fig_TPU}), it is possible to generalize the operation also to negative and complex input vectors. 
Note that here all the matrix multiplications are performed in parallel in one step. The parallelism can be further improved by using multiple MRR resonance orders at the same time. In this case, every MRR will weight all the input wavelengths that couple to its resonances, and these will be then divided by WDM filters before being detected. The operation time-of-flight latency is here only limited by the detection speed, which can be lowered to tens of picoseconds. Note also that in machine learning tasks, once concluded the training phase, the weights are usually fixed or slowly updated in time, with respect to the input vector. In this situation, the device becomes particularly efficient, as once the weights are loaded by tuning the heaters of each MRR, the structure is able to process a fast input signal requiring only a constant energy power to keep the MRR transmissions at the trained value. Additionally, the energy efficiency can be further improved by adopting phase change materials (PCMs), as they allow for nonvolatile photonics memory that only demands energy when the weight needs to be updated. This solution has been investigated with PCMs based on $\rm Ge_2Sb_2Se_2$ patches, $30$ nm thin and $250$ nm wide, placed on top of the waveguide and arranged in a grating fashion, within a structure closely similar to the MRR-based weight bank \cite{miscuglio2020photonic}. By local electrostatic heating, the PCMs can be individually and reversibly switched between an amorphous and crystalline phase, characterized by different absorption coefficients. By simulation, it was demonstrated that by cascading 15 PCM stages a memory with total insertion losses limited to only $1\,\rm dB$ (all PCMs in the amorphous state), extinction ratio up to $3.5\, \rm dB$ (all PCMs in the crystalline state) and 4-bit resolution can be achieved. Alternatively, a 4-bit resolution can be obtained by only 4 PCMs having different lengths. While this option positively reduces the number of heating wires required, it may demand on the other hand different phase-loading times and higher voltages to change the PCM phases.

Matrices of MRRs have been also proposed as the input neurons of a convolutional neural network in \cite{wang2023photonic}. The MNIST dataset \cite{MNIST} is encoded in the optical domain by means of matrices of MRRs. The original image is first transformed into a binary image. Then, its pixel values are codified in the wavelength detuning of an equal number of MRRs with respect to the input signal wavelengths. Pixel values one and zero correspond to conditions where the MRRs are in resonance, or out of resonance, with respect to the input wavelength. Then the coded optical image is  fed to a convolutional neural network based on a cross-bar network of MRRs. Binary weights to the kernel are applied via MRRs detuning with respect to the input wavelengths. Output data are collected by the nonlinear response of photodetectors at the convolutional matrix output. High classification rates with a very high energy efficiency are demonstrated.

\section{Hybrid approaches to silicon-based neural networks}
\label{sec:Hybrid}
The properties and functionalities of silicon MRRs can be enhanced and expanded by specific materials added to the integration process (such as PCMs), resulting in hybrid silicon photonic devices. In turn, MRRs can also improve and facilitate the use of these materials, thanks to their strong sensitivity to wavelength and optical path perturbations, and to their power enhancement capability. In this section we summarize some of the most relevant works proposing photonic artificial neurons or activation functions based on hybrid MRRs.

\begin{figure}[b!]
	\centering\includegraphics[scale=0.50]{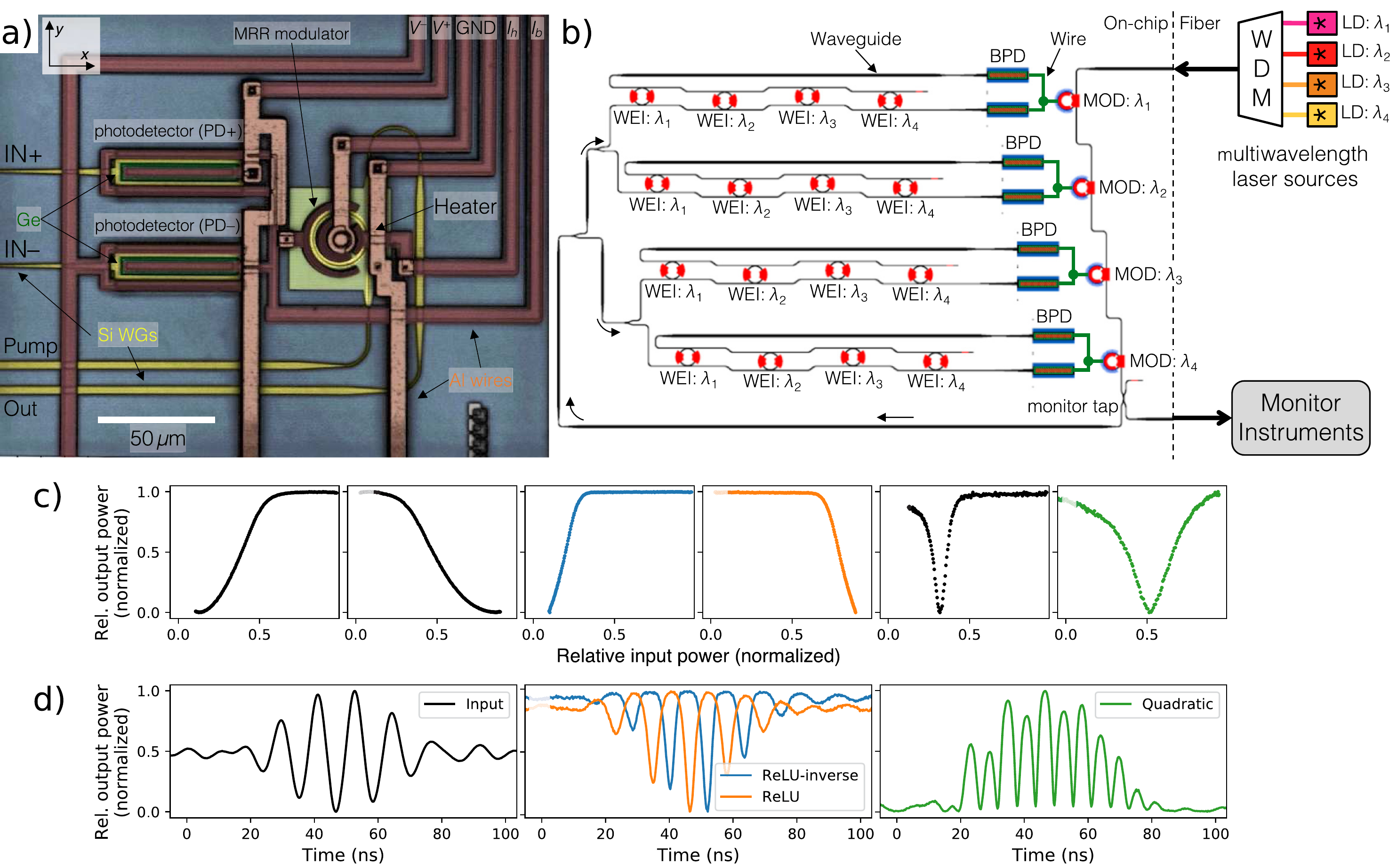}
	\caption{a) Micrograph of the integrated photonic neuron discussed in \cite{prucnalSiliconNeuron2019}, consisting of a silicon MRR modulator, actuated by the output of a balanced photodetector. b) Schematic of a proposed broadcast-and-weight integrated photonic ANN, featuring the presented photonic modulator neuron. c) Experimentally obtained activation functions performed by the photonic neuron, obtained by inserting a slow-rising saw-tooth input waveform at 200 kHz. Depending on the setting of the driving photodetectors, different transfer functions of the MRR modulator can be obtained. In this way, different activation function for the photonic neurons are generated, which spans from sigmoids (first two panels on the left), to ReLU (two central panels), to radial basis or quadratic functions (two right panels). d) 40-ns burst of a 100-MHz carrier employed as input signal (left plot) and the corresponding MRR time resolved transmission when the MRR is set to implement the activation functions indicated in the legend. Reprint by permission from \cite{prucnalSiliconNeuron2019}.}
	\label{fig1_hybrid}
\end{figure}

In \cite{prucnalSiliconNeuron2019}, a silicon MRR is doped  to form a pn junction for fast optical modulation via free carriers injection. The modulator driving current is set by an integrated balanced photodetector (see Fig. \ref{fig1_hybrid} (a)) while the MRR resonance can be adjusted by an n-type doped heater. This artificial neuron performs optical-to-electrical-to-optical conversion, in order to process multi-wavelength optical inputs and outputs, and to manifest configurable nonlinear activation functions. The balanced photodetector accepts two optical inputs, one excitatory (positive current) and the other inhibitory (negative current). The two photodetector inputs induce opposite MRR modulations which are also influenced by the MRR set point due to the heater. In this way, different MRR transfer functions can be obtained (Fig. \ref{fig1_hybrid} (c)), which impact on the MRR time response to a oscillating carrier signal (Fig. \ref{fig1_hybrid} (d)).
The proposed artificial neuron is designed to be integrable with the photonic network presented in \cite{tait2017neuromorphic}, which is discussed in Section \ref{sec:weight_bank}. Figure \ref{fig1_hybrid} (b) shows the schematics of the fully integrated photonic ANN, where silicon MRRs are used to implement both synaptic weights (via heaters) and artificial neurons (with balanced photodetector and \textit{p-n} modulation). In particular, this kind of photonic neuron demonstrates nonlinear activation function configurability (see Fig. \ref{fig1_hybrid} (c) and (d)), fan-in (i.e. capability of combining multiple independently weighted inputs) and cascadability. Moreover, this kind of artificial neuron has also pulse compression capability (i.e. limiting broadening pulses propagating through the network). 

\begin{figure}[t!]
\centering\includegraphics[scale=0.25]{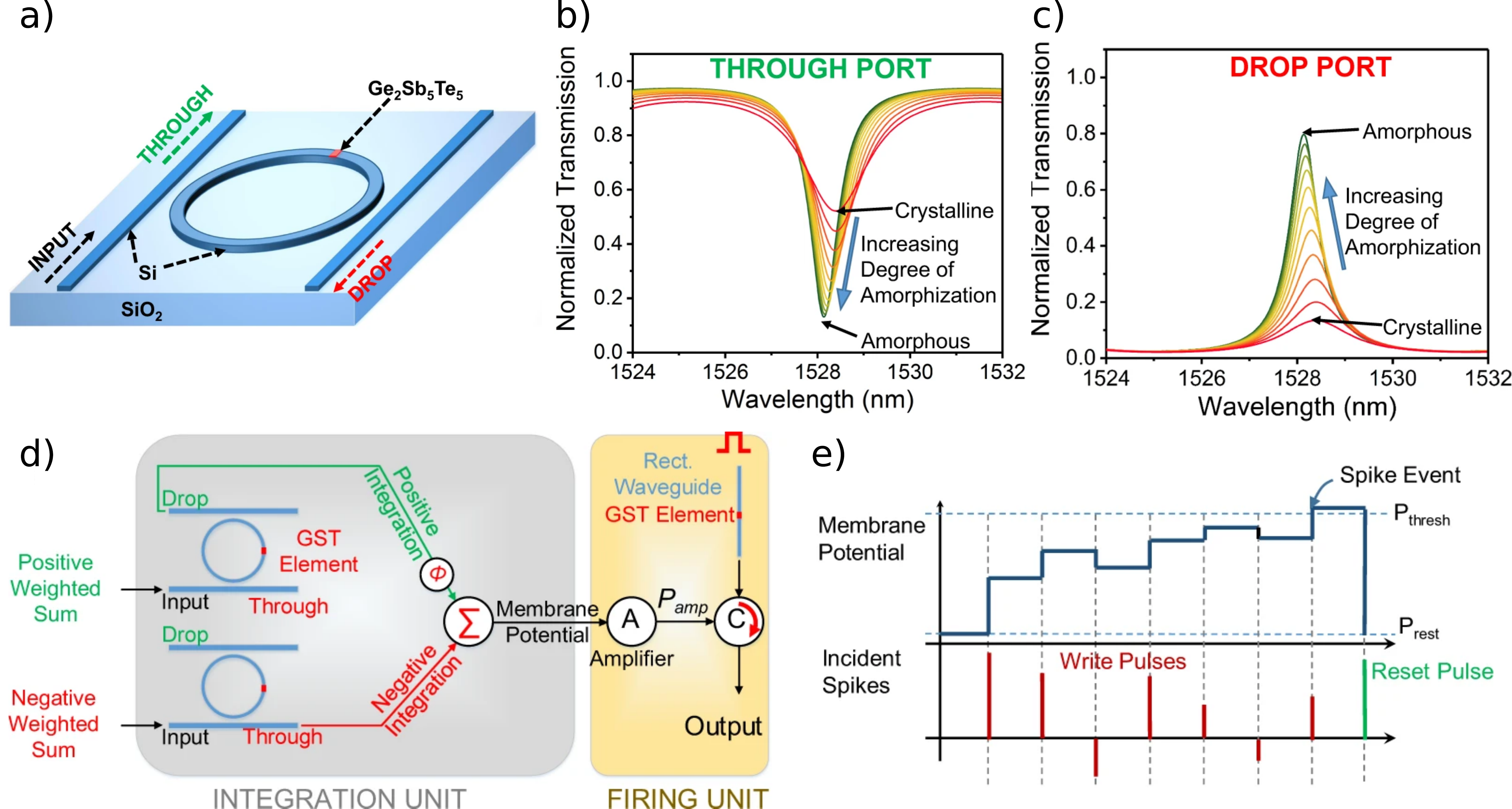}
\caption{a) Schematic of a silicon MRR with PCM element, considered in \cite{chakraborty2018toward} to integrate neuron inputs over time. b) and c) MRR spectrum for different levels of GST amorphization, respectively at the through and the drop port. d) Integrate-and-fire spiking neuron based on MRRs and PCM. e) Diagram of the neuron internal state as a function of time, for different subsequent input pulses. Reprint by permission from \cite{chakraborty2018toward}.}
\label{fig2_hybrid}
\end{figure}

In \cite{chakraborty2018toward}, a simulation work shows how two MRRs can be combined with a phase change material (namely GST, short for the Ge\(_2\)Sb\(_2\)Te\(_5\) alloy) to build an integrate-and-fire spiking neuron. In particular, the neuron functionality is based on the resonance change in MRRs due to different non-volatile states of a PCM cell, which consists of a thin GST layer deposited on a small area (300\(\times\)300 nm\(^2\)) on top of the ring waveguide (see Fig. \ref{fig2_hybrid} (a)). When the GST is in a fully crystalline state (with high absorption), most of the resonant optical input power pasts the MRR, along the through port, while only a minor part is redirected to the drop port (see Fig.s \ref{fig2_hybrid} (b), (c)). The higher the amorphization level of the GST cell (the lower its optical absorption),  the higher the power sent to the drop port and the lower the power at the through port. The GST amorphization level is changed all-optically, by means of 200 ps long pulses, ranging from 12 mW to 26 mW. 

This property can be used to integrate input pulses over time. In particular, the integration unit is based on two MRRs, whose inputs are the output pulses from the previous neurons with positive and negative synaptic weights respectively (Fig. \ref{fig2_hybrid} (d)). Then, the combination of the drop output of the first MRR and of the through output of the second MRR, adjusting the phase shift between the two, is performed. The resulting pulse is then amplified and reaches a \textit{firing} waveguide with a GST element in crystalline state. Once the combined amorphization level of the GST at the two MRRs (and thus the sum of the weighted inputs over time) reaches a threshold value (Fig. \ref{fig2_hybrid} (e)), the GST element is amorphized and a pulse, independently generated and inserted into the firing waveguide, is transmitted to the output of the firing neuron. The GST states need to be reset by suitable optical pulses after each neuron activation.
Finally, it is shown that a simulated 3-layers feed-forward spiking ANN comprising the proposed neuron devices reaches 98.06\% accuracy on the MNIST classification task \cite{MNIST}, with 0.24\% degradation with respect to an ideal version of the network.

\begin{figure}[t!]
\centering\includegraphics[scale=0.25]{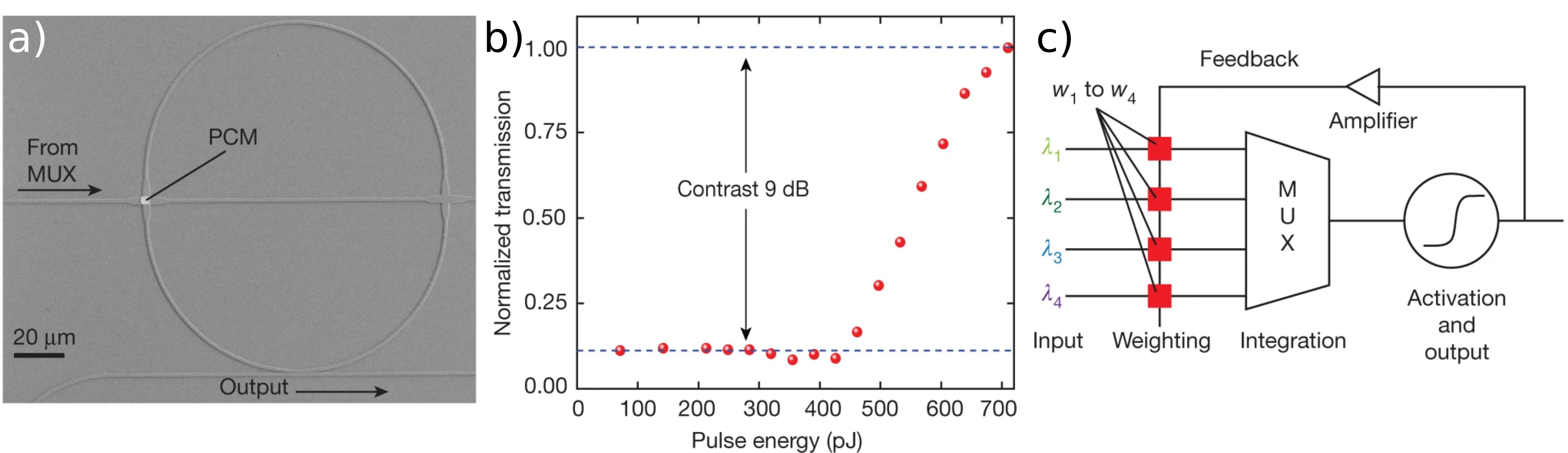}
\caption{a) Micrograph of the photonic neuron based on a MRR with PCM cell (GST) presented in \cite{feldmann2019all}. Input pulses are inserted through the upper straight waveguide, while the lower one is used to transmit output pulses, depending on the amorphization level of the PCM cell. b) Activation function of the optical neuron, showing the transmission of the output waveguide coupled to the MRR, as a function of the input pulse energy. c) Schematic of the circuit performing the weighting of the neuron inputs and the feedback connection employed to implement synaptic plasticity. Reprint by permission from \cite{feldmann2019all}.}
\label{fig3_hybrid}
\end{figure}

A similar but simpler approach is experimentally demonstrated in \cite{feldmann2019all}, where a full all-optical spiking network is presented. In this case, the neuron consists of a MRR coupled to a single straight waveguide (all-pass configuration) and crossed by another straight waveguide (Fig. \ref{fig3_hybrid} (a)), where input pulses of different wavelengths (wavelength division multiplexing) are inserted. A GST layer (9 \text{\(\mu\)}m\(^2\)) is deposited on top of one crossing so that input pulses can permanently modify the MRR transmission by all-optically changing the GST amorphization level. In this way, the combined strength of the weighted input pulses determines the achieved GST amorphization level. Independently generated pulses are sent through the coupled waveguide (at the bottom of Fig. \ref{fig3_hybrid} (a)) and they are transmitted only if the GST at the crossing is sufficiently amorphized. Therefore, the output pulse transmission is a nonlinear function (i.e. the activation function of the neuron) of the overall energy delivered by the input pulses (Fig. \ref{fig3_hybrid}(b)). With a further connection between the neuron output and the input synaptic weights (implemented using GST cells on straight waveguides, see Fig. \ref{fig3_hybrid} (c)), synaptic plasticity is introduced. This is based on the strengthening of those input connections that deliver pulses strong enough to make the neuron fire. Such an effect, often summarized with the sentence `neurons that fire together, wire together', is known to be one of the key mechanisms allowing the animal brain to memorize and learn (Hebbian learning) \cite{lowel1992selection}.
The maximum pulse energy employed to modify the GST cell solid state is 710 pJ, delivered by 200 ns optical pulses.
An all-optical single-layer ANN based on these building blocks, which solves simple pattern recognition tasks, is also demonstrated.

\begin{figure}[t!]
\centering\includegraphics[scale=0.18]{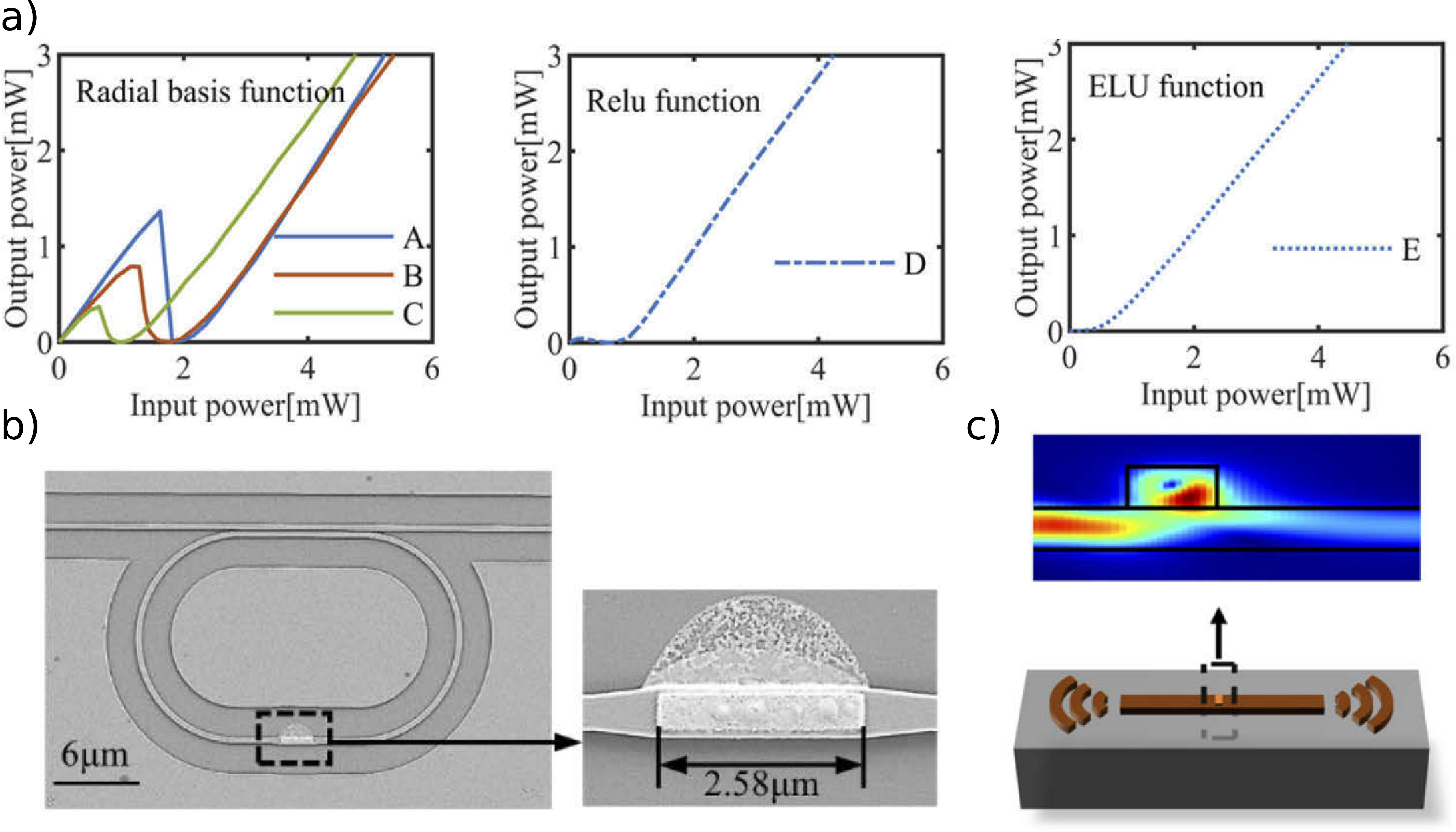}
\caption{a) Experimental activation functions demonstrated in \cite{LowThresholdGeSi2022} by means of: b) a silicon racetrack MRR with Ge layer (shown in the zoomed in image). c) Simulated germanium resonant structure on top of a Si straight waveguide. The upper image shows the field profile along the waveguide and the Ge structure. Reprint by permission from \cite{LowThresholdGeSi2022}.}
\label{fig4_hybrid}
\end{figure}

In \cite{LowThresholdGeSi2022}, the authors experimentally demonstrate three types of all-optical activation functions (Radial basis, ReLU and ELU, see Fig. \ref{fig4_hybrid} (a)) employing a silicon MRR whose waveguide is partially covered by a layer of deposited germanium (2.58 \(\mu\)m long, see Fig. \ref{fig4_hybrid} (b)). This is achieved by exploiting the thermo-optic effect of germanium in conjunction with the silicon MRR properties, considering a constant and coherent optical input. The different shapes of the activation function are obtained by employing different input wavelengths, considering the resonance wavelength of the MRR as a reference.
The device is CMOS-compatible and it presents a  nonlinear threshold down to 0.74 mW in input optical power (on-chip), working with a repetition rate below 100 kHz. The authors also show a simulated germanium nanostructure (a sub-\(\mu\)m block on top of a straight Si waveguide, see Fig. \ref{fig4_hybrid} (c)) that could host an optical resonance, thus providing a nonlinear activation function within a smaller footprint while working at a higher speed. Finally, the usability of these activation functions is validated in a simulated ANN to tackle the MNIST classification task \cite{MNIST}.

\begin{figure}[t!]
\centering\includegraphics[scale=0.20]{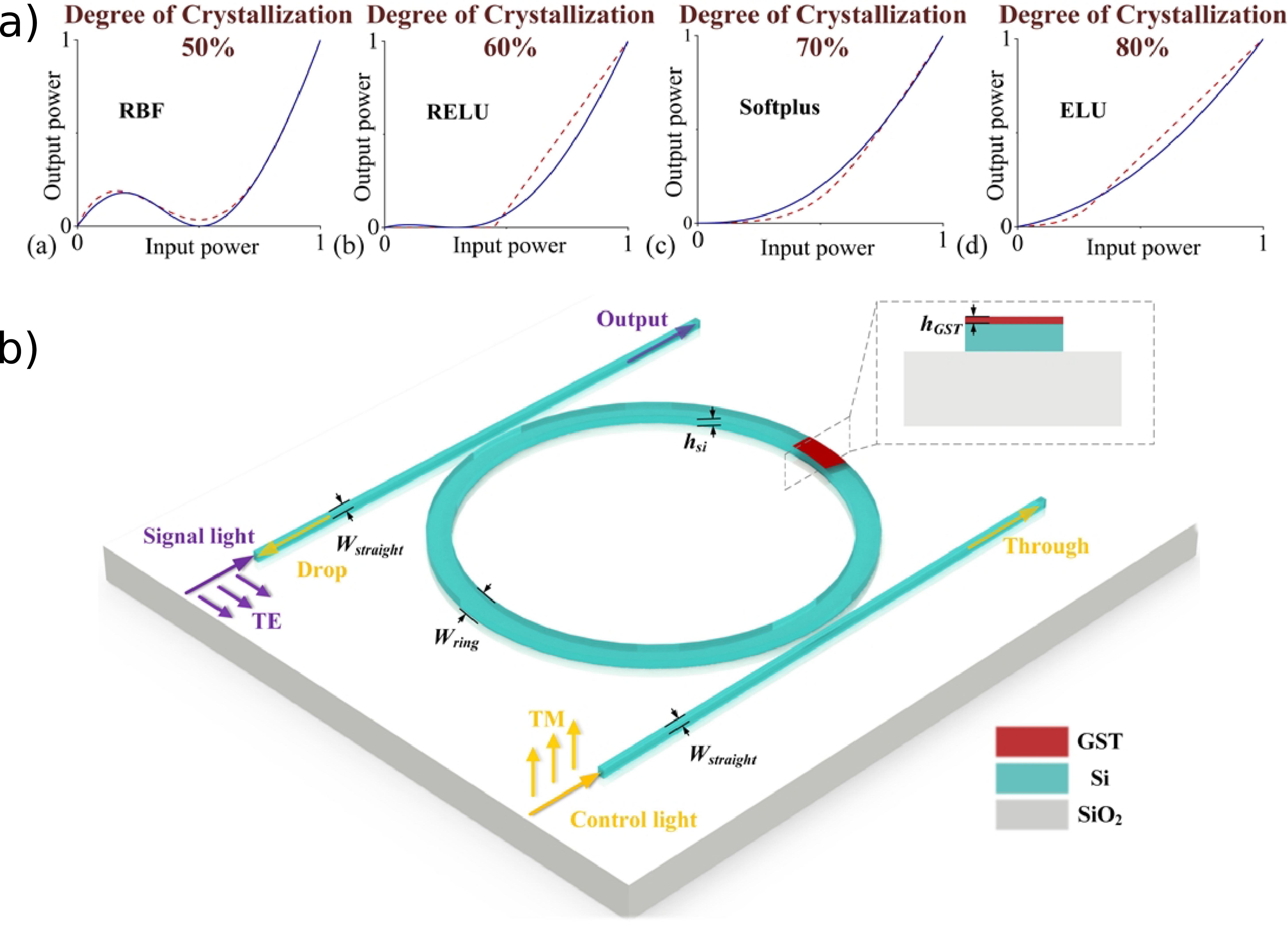}
\caption{a) Different activation functions corresponding to different degrees of amorphization of the PCM layer deposited on the silicon MRR. Black lines refer to the MRR transfer function while the dashed red lines to the ideal activation function. b) Schematic of the simulated device, i.e. an add-drop MRR where part of the ring waveguide is covered with GST. A constant signal with TE polarization is used as input to the activation function. A TM optical pulse is inserted in the parallel port to switch the GST and thus to change activation function. Reprint by permission from \cite{ProgLowPowerPhaseChange2022}.}
\label{fig5_hybrid}
\end{figure}

In \cite{ProgLowPowerPhaseChange2022}, simulations show that four different types of all-optical activation functions (approximating the ReLU, ELU, Softplus and radial basis function, see Fig. \ref{fig5_hybrid} (a)) can be achieved by setting different amorphization levels in a short GST layer (0.5 \(\mu\)m long), deposited on a silicon add-drop MRR (see Fig. \ref{fig5_hybrid} (b)). Differently from the previously discussed GST-based nodes \cite{chakraborty2018toward, feldmann2019all}, the nonlinear response (to a constant optical input) is provided by the silicon nonlinear effects (due to free-carriers and heating), and not by switching the PCM. The GST amorphization level is indeed kept constant during the device activity (less than -3 dBm reaches the waveguide segment covered by GST) and it is switched only to program the activation function shape. This is done by inserting optical pulses with TM (transverse magnetic) polarization (Fig. \ref{fig5_hybrid} (b)), so that the field along the ring waveguide overlaps better with the GST layer, increasing the energy efficiency. To amorphize the PCM, a 1 ns pulse of 107 mW power was considered, while a double-step pulse was used for crystallization. The first part of such a pulse is the same as the amorphizing pulse, the second part lasts 29 ns and has 10 mW power. The achieved activation functions were employed in a simulated ANN to classify images from the MNIST dataset \cite{MNIST}, obtaining comparable performances to when the \textit{tanh} activation function is employed.

\begin{figure}[t!]
\centering\includegraphics[scale=0.25]{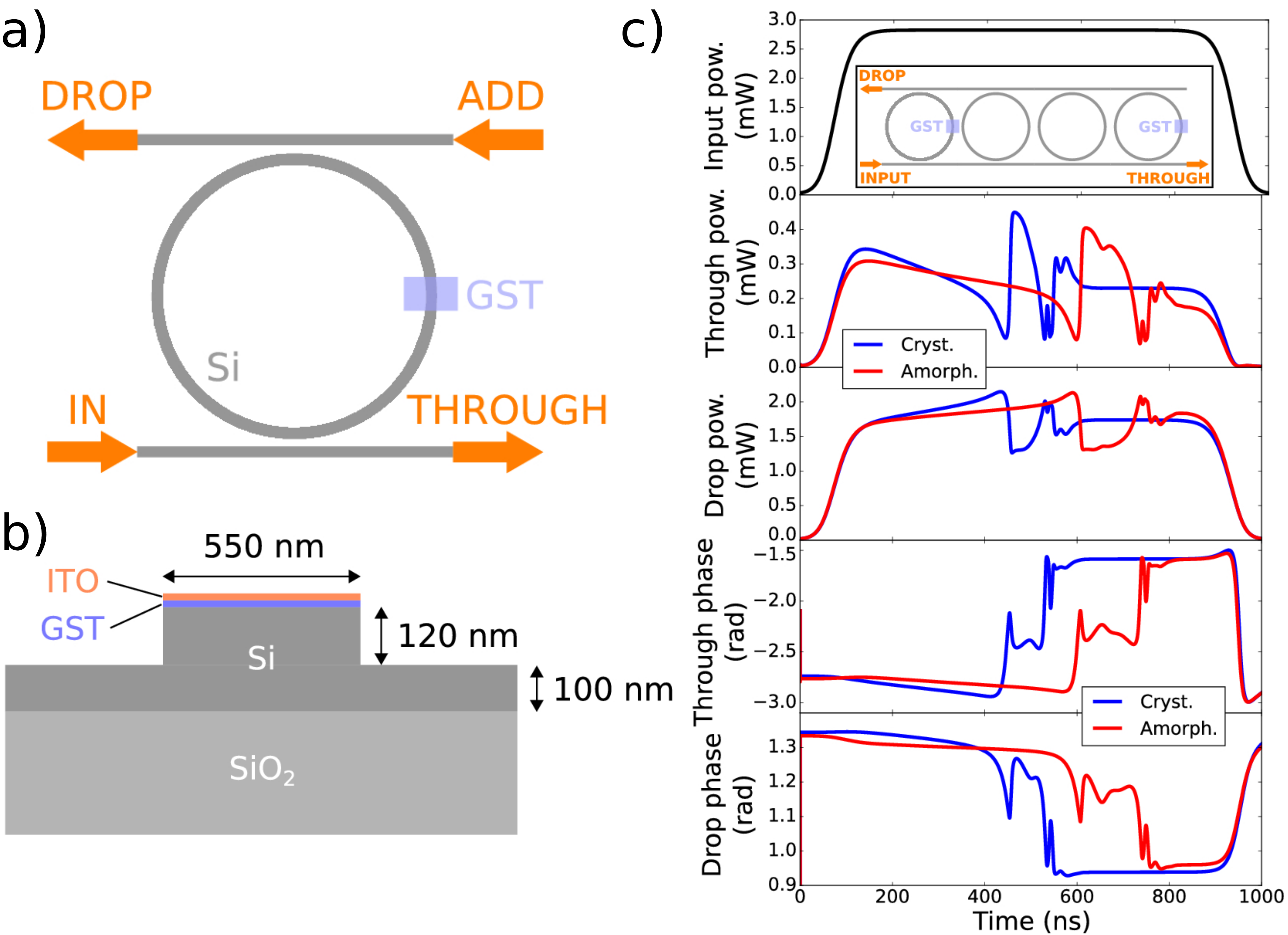}
\caption{a) Schematic of the add-drop silicon MRR proposed in \cite{lugnan2022rigorous} as a plastic neuromorphic node. b) Cross section of the simulated GST cell, included in the MRR model. c) Input and output plots of the simulated silicon photonic circuit schematized in the inset of the top plot. Blue and red data respectively show the response for minimum and maximum GST amorphization levels. From top to bottom, the plots show input optical power, output power and optical phase at through and drop ports. Reprint by permission from \cite{lugnan2022rigorous}.}
\label{fig6_hybrid}
\end{figure}

In \cite{lugnan2022rigorous}, a more rigorous numerical model of a similar structure is discussed. Here, an add-drop silicon MRR with a short GST patch on top of part of the ring waveguide (see Fig. \ref{fig6_hybrid} (a) and (b)) is modeled. The aim is to use this structure as a dynamic neuromorphic node with synaptic plasticity, to build scalable and plastic photonic ANNs. The presented equations (based on the time-dependent coupled mode theory) model the main nonlinear effects in silicon (due to free carriers and heating) together with the additional thermo-optic effect at the waveguide segment covered by GTS, and the PCM amorphization level. The model accounts for the asymmetry and the non-idealities due to the high optical loss introduced by GST and the consequent large MRR coupling coefficients (so as to stay close to the critical coupling conditions). Simulations show that speed and energy efficiency of all-optical memory operations (i.e. switching and resetting the GST via optical pulses) are considerably improved with respect to a simple straight silicon waveguide with GST. These advantages are due to the resonance of the MRR, namely the power enhancement and the increased sensitivity to complex refractive index perturbations along the ring waveguide. In particular, a resettable 15\% contrast was demonstrated for a 0.7 $\mu$m long GST cell, employing a 10 ns optical pulse conveying around 0.26 nJ for amorphization, and  a 4-pulses series with 70 ns overall duration, conveying less than 0.5 nJ, for recrystallization. Finally, a small network of four coupled MRRs (see Fig. \ref{fig6_hybrid} (c)), two with PCM and two without, shows high contrast in the dynamical response for two different memory states of the GST cells. In particular, the self-pulsing behaviour of the coupled MRRs significantly depends on the GST non-volatile state.

\section{Conclusions and perspectives}\label{sec:Conclusions}

In this review, we have discussed several different implementations of neural networks building blocks based on microresonators. A summary of the advantages and limitations of microresonators for application in neural networks is provided in Table 1. Currently, experimental studies using microresonators have been limited to simple arrays with only a few resonators. However, it would be intriguing to explore the use of microresonators in more complex arrangements to enlarge the phase space in which input data are projected. In this context, the use of small worlds neural networks is particularly appealing. In the context of neural networks, small worlds refer to complex geometric arrangements of interconnected nodes, where some nodes are more strongly connected than others. Clustering and long-range random connections between nodes increase the network representation of complex problems, allowing for more efficient and effective processing. By incorporating microresonators into these types of arrangements, it may be possible to create more complex and powerful photonic neural networks which use both the long and short-term memories of microresonators as well as the multiple inputs and outputs of array of microresonators. In particular, configurations that include both short-range and long-range connections may be especially promising. These types of configurations could potentially mimic other brain functionalities, as the brain is known to contain both local and long-range connections between neurons.
\begin{table}[tp!]
	\caption{The advantages and disadvantages of the use of microring resonators as optical neurons in artificial neural network}
    \centering
    \begin{tabularx}{\linewidth}{L{0.53\linewidth} L{0.43\linewidth}}
		\hline
		\thead{Pros}                      & \thead{Cons}                \\ \hline 
		\begin{itemize}
        \setlength\itemsep{0em}
        \justifying
        \item Tuned with low energy (either when thermally, electrically or optically operated)
        \item Nonlinear optical properties at low-power due to the field enhancement in the cavity
        \item Based on a resonant cavity that enhances the dependence on external parameters
        \item High sensitivity to perturbations	
        \item Resonances are periodic in frequency and almost in wavelength. \item Free spectral range is adjusted by design
        \item Different temporal dynamics and lifetimes for different nonlinear effects
        \item Different capability to store information at different time scales which imply a fast, fading and slow memory depending on the involved non-linearity
        \item Transfer function (relation between input and output) can emulate many different neuronal activation functions and is externally tunable
        \item High dynamic range in the output with a large extinction ratio between ``on" and ``off" states
        \item Stationary response and pulsed response (self-pulsing, spiking) to a CW input
        \item Can integrate different input temporal pulses or different input wavelengths
        \item Information can be processed either if it is carried by the phase  of the input optical signal or by the amplitude or by both
        \item The connection between different MRRs can be active or passive (the state of one can be independent of or be affected by the state of the following, including the MRR resonance position)
        \item Small size and compact even in large numbers
        \item Ease of cascadability (arranged in sequences or matrices)	with different connection schemes (direct such as in CROW or indirect such as in SCISSOR)
        \item Based on a mature fabrication technology (silicon photonics)
        
        \end{itemize} &
        \begin{itemize}
        \setlength\itemsep{0em}
        \justifying
        \item Optical energy is dissipated in the MRR-based ANN, which limits the ANN sizes due to losses 
        \item High insertion losses
        \item The MRR resonance randomly depends on the fabrication errors and is thermally unstable 
        \item Other MRR properties (e.g. Q-factor, free spectral range, FWHM, coupling coefficient) strongly depend on the fabrication (need a post-fabrication active trimming)
        \item Silicon MRRs are passive optical devices, no gain unless heterointerfaced with other materials
        \item Limited number of input and output channels. Increasing the number of channels induces losses and reduces the Q-factor. 
        \item Thermally tuning of MRRs causes both local and global thermal cross-talk across the chip         
        \item Difficult to use negative weights if data are encoded in the optical signal intensity
        \end{itemize} \\
	   \hline
	\end{tabularx}
\end{table}

One aspect where PICs currently fall short is in their ability to create multilayer structures with intralayer connections of tunable strength. This limitation prevents from creating architectures that mimic the intricate three-dimensional structure of the brain, where different regions can influence each other even when far apart. By incorporating microresonators into such multilayer structures, it may be possible to create more realistic models of the brain's neural networks and improve our understanding of brain function. Here, three-dimensional integration of photonic components is needed. 

To enhance the performance of microresonators as neural nodes, researchers are exploring the integration of different materials with unique properties. Examples are phase-change materials, which have been successfully integrated with microresonators in previous studies. In addition, the use of lithium niobate (LiN) is also promising due to its fast modulation capabilities and nonlinear properties. LiN allows for fast switching and new frequency generation, which can enable more efficient and effective processing in photonic neural networks. Other materials that can be integrated with microresonators are direct gap semiconductors. These offer characteristics such as electro-absorption, efficient and fast detectors, and optical amplification, which can significantly enhance the performance of microresonators as neural nodes. By combining the unique properties of these different materials with microresonators, it may be possible to create more advanced photonic neural networks where optical losses are no longer limiting the ANN size and complexity.

Silicon photonics technology enables seamless integration of photonics and electronics, providing an opportunity to leverage the strengths of both domains for developing more efficient and powerful artificial neural networks. By integrating different types of active nodes with varying properties such as activation function, temporal response, and interconnection geometry, we can significantly enhance the performance of artificial neural networks. The goal is to create neural networks that not only have neurons, but also other types of cells, such as glial cells, that are present in the human brain. The use of silicon technology allows for the creation of highly compact and densely packed integrated circuits that can process large amounts of data at high speeds. With the integration of different types of neurons and glial cells, artificial neural networks can be made more biologically realistic and perform a wider range of tasks. For example, incorporating inhibitory neurons into the network can help prevent over-excitation, which can lead to instability. Furthermore, incorporating glial cells, which play a crucial role in the brain's immune system, can help protect the network from external threats and ensure its proper functioning.

By having a variety of different types of active nodes with distinct properties that are weakly or strongly interconnected, we can significantly enhance the performance of ANNs. This can enable the network to perform tasks that are currently beyond the capabilities of traditional computing systems. Overall, the integration of silicon photonics technology with different types of active nodes holds tremendous promise for the development of advanced ANN that can perform complex tasks with high accuracy and efficiency.

\section*{Acknowledgments}
We gratefully thank Apostolos Argyris, Davide Bazzanella, Paolo Bettotti, Riccardo Franchi and Claudio Mirasso for useful suggestions and interesting discussions. 

\subsection*{Author Contributions}
S. Biasi wrote section 2, S. Biasi and G. Donati wrote section 3, G. Donati wrote section 4, A. Lugnan and M. Mancinelli wrote section 5, L. Pavesi wrote sections 1 and 6. All the authors finalized the manuscript.

\subsection*{Funding}
This research was supported by the European Research Council (ERC) under the European Union's Horizon 2020 research and innovation program [grant agreements No 788793, BACKUP, and No 963463, ALPI] and from the MUR under the project PRIN PELM [grant number 20177 PSCKT]. S. Biasi acknowledges the co-financing of the European Union FSE-REACT-EU, PON Research and Innovation 2014–2020 DM1062/2021. A. Lugnan acknowledges the funding by the European Union under GA n°101064322-ARIADNE.

\subsection*{Conflicts of Interest}
The authors declare that there is no conflict of interest regarding the publication of this article.

\subsection*{Data Availability}
The data that support the findings of this study are available from the corresponding author upon reasonable request.

\bibliographystyle{unsrt}
\bibliography{references}  






\end{document}